\documentclass[pra,aps,twocolumn,groupedaddress,showpacs,showkeys,superscriptaddress]{revtex4-2}
\usepackage{amsmath}
\usepackage{eurosym}
\usepackage{amssymb}
\usepackage{graphicx}
\usepackage{dcolumn}
\usepackage{bbm}
\usepackage{epsfig}
\usepackage{xcolor}
\usepackage[breaklinks,colorlinks,bookmarks=false,citecolor=blue,linkcolor=red,urlcolor=blue]{hyperref}
\usepackage[title]{appendix}

\begin{document}

\title{Enhancing fidelity in teleportation of a two-qubit state via a quantum communication channel formed by spin-1/2 Ising-Heisenberg trimer chains due to a magnetic field}
\author{Jozef Stre\v{c}ka}
\email{Corresponding author: jozef.strecka@upjs.sk}
\affiliation{Department of Theoretical Physics and Astrophysics, Faculty of Science, P.~J. \v{S}af\'arik University, Park Angelinum 9, 04001 Ko\v{s}ice, Slovakia}
\author{Fadwa Benabdallah}
\affiliation{LPHE-MS, Department of Physics, Faculty of Sciences, Mohammed V University in Rabat, Morocco}
\author{Mohammed Daoud}
\affiliation{LPMS, Department of Physics, Faculty of Sciences, University Ibn Tofail, Kenitra, Morocco}
\affiliation{Abdus Salam International Centre for Theoretical Physics, Miramare, Trieste, Italy}

\begin{abstract}
We demonstrate that two independent spin-1/2 Ising-Heisenberg trimer chains provide an effective platform for the quantum teleportation of any entangled two-qubit state through the quantum communication channel formed by two Heisenberg dimers. The reliability of this quantum channel is assessed by comparing the concurrences, which quantify a strength of the bipartite entanglement of the initial input state and the readout output state. Additionally, we rigorously calculate quantities fidelity and average fidelity to evaluate the quality of the teleportation protocol depending on temperature and magnetic field. It is evidenced that the efficiency of quantum teleportation of arbitrary entangled two-qubit state through this quantum communication channel can be significantly enhanced by moderate magnetic fields. This enhancement can be attributed to the magnetic-field-driven transition from a quantum antiferromagnetic phase to a quantum ferrimagnetic phase, which supports realization of a fully entangled quantum channel suitable for efficient quantum teleportation. The polymeric trimer chains Cu$_3$(P$_2$O$_6$OH)$_2$ are proposed as an experimental resource of this quantum communication channel, which provides an efficient platform for realization of the quantum teleportation up to moderate temperatures 40~K and extremely high magnetic fields 80~T.
\end{abstract}

\maketitle
\section{Introduction}

Quantum-mechanical systems capable of serving as resources for quantum information processing are among the most sought-after physical systems today due to their significant application potential in emerging quantum technologies \cite{jaeg18}. Over the past few decades, considerable effort has been devoted to exploring and quantifying entanglement in condensed matter systems \cite{AMICO2008, HORODECKI2009, GUHNE2009}. Entanglement has emerged as a fundamental resource, which is indispensable for advancing quantum technologies, particularly in the domains of quantum communication, quantum information processing, and quantum computation \cite{vinc05,niel10,ladd10}. In this context, solid-state magnetic materials affording the experimental realization of quantum spin systems play a pivotal role in entanglement research \cite{bene15,stre23,cruz23,pint23,pine24}. Among these, one-dimensional quantum Heisenberg spin chains stand out as some of the simplest quantum spin systems in which entanglement naturally arises \cite{CONNOR2001,KAMTA2002,SUN2003}. 

In recent years, quantum teleportation has become a key process in the rapidly advancing field of quantum information science. Since Bennett's groundbreaking proposal \cite{BENNETT1993}, the quantum teleportation has been extensively explored in both theoretical \cite{POPESCU1994,HORODECKI1999,YEO2002,rigo2005,ZHANG2007,KH2008,ZHOU2008,QIN2015,rigo2017,rigo2023,rigo2024} as well as experimental \cite{BOUWMEESTER1997,BOSCHI1998} contexts, as it stands out as one of the most fascinating and counterintuitive demonstrations of nonlocal quantum properties. Among various quantum teleportation schemes, those utilizing  antuferromagnetic spin-1/2 Heisenberg chains have proven particularly effective and efficient as communication channels with the intriguing added capability of detecting quantum critical points \cite{rigo2023,rigo2024}. Recent studies have also investigated quantum teleportation facilitated by quantum channels composed of two-qubit pairs embedded within exactly solvable Ising–-Heisenberg spin chains such as a diamond spin chain \cite{ROJAS2017,ROJAS2019} or a branched spin chain \cite{ZHENG2019,ZAD2021}.

The spin-1/2 Heisenberg trimer chain represents historically a first example of a quantum spin chain, for which a fractional magnetization plateau of purely quantum origin was theoretically predicted \cite{HIDA1}. This model gained further attention particularly after the detection of an intermediate one-third magnetization plateau in the copper-based polymeric compound Cu$_{3}$(P$_{2}$O$_{6}$OH)$_{2}$ affording its experimental realization \cite{HASE1,HASE2,HASE3,KONG2015,HASE4}. Due to its complexity, the magnetization curves and low-temperature behavior of the spin-1/2 Heisenberg trimer chain have been extensively studied using various numerical techniques as for instance the exact diagonalization \cite{HIDA2}, the density-matrix renormalization group method \cite{SHU2008}, the transfer-matrix renormalization group technique \cite{GU2006}, and the strong-coupling expansion \cite{HONECKER1999}. A modified strong-coupling approach developed for the spin-1/2 Heisenberg trimer chain from an exactly solved Ising-Heisenberg diamond chain enabled refinement of the coupling constants for the polymeric trimer chain Cu$_{3}$(P$_{2}$O$_{6}$OH)$_{2}$ by matching theoretical predictions with available experimental data for magnetization and magnetic susceptibility in a wide range of temperatures and magnetic fields \cite{VER21}. Besides, studies using quantum Monte Carlo, exact diagonalization, and perturbation theory revealed intriguing doublon and quarton quantum excitations in the spin-1/2 Heisenberg trimer chain \cite{cheng22,cheng24}. These findings were experimentally confirmed by recent inelastic neutron scattering measurements on the copper-based trimeric chain Na$_2$Cu$_3$Ge$_4$O$_{12}$ \cite{bera22}. The topologically nontrivial nature of the one-third magnetization plateau and fractional excitations was also independently verified through high-field nuclear magnetic resonance measurements on the trimeric compound Na$_2$Cu$_3$Ge$_4$O$_{12}$ and its silicon-doped analogues Na$_2$Cu$_3$Ge$_{4-x}$Si$_{x}$O$_{12}$ \cite{han24}.

Despite considerable efforts, exact results are available only for specific simplified versions of trimeric spin chains such as the spin-1/2 XY trimer chain \cite{OKAMATO1996,ZVYAGIN1991,DING2010} and the spin-1/2 Ising-Heisenberg trimer chain \cite{JO2003,JO2017}. The ground state, magnetization process, and thermodynamic properties of the exactly solved spin-1/2 Ising-Heisenberg trimer chain were thoroughly investigated in Ref. \cite{JO2003}, whereas the bipartite entanglement and non-locality in the spin-1/2 Ising-Heisenberg trimer chain were later confirmed  at both zero as well as non-zero temperatures in Ref. \cite{JO2017}. Motivated by these developments, this work aims to investigate quantum teleportation of an arbitrary entangled two-qubit state using a quantum channel formed by two independent spin-1/2 Ising-Heisenberg trimerized chains. Specifically, we investigate the influence of temperature, magnetic field, and coupling constants on the key characteristics including the entanglement of the quantum channel and the output state in addition to the fidelity and average fidelity of the quantum teleportation protocol. Our findings provide compelling evidence for the high reliability of the this quantum channel making it a promising candidate for efficient quantum communication. While prior studies on quantum teleportation have primarily focused on teleporting perfectly entangled Bell state \cite{BEN2022,BEN12022}, this study demonstrates that partially entangled pure states can be also transmitted through the quantum channel with remarkably high fidelity.

The structure of this paper is as follows. In Sec. \ref{model}, we introduce the spin-1/2 Ising-Heisenberg trimer chain and outline the key steps for deriving its reduced (two-qubit) density operator using the transfer-matrix formalism. In Sec. \ref{QTF}, we present a detailed quantum teleportation scheme including a rigorous calculation of the concurrence for the quantum channel, input and output states, as well as, the fidelity and average fidelity of the teleportation protocol. Sec. \ref{Results} presents  the most significant theoretical finding obtained for this quantum communication channel accompanied by an in-depth discussion. Sec. \ref{resexp} explores the potential for practical realization of quantum teleportation via a communication channel formed by two polymeric trimer chains Cu$_{3}$(P$_{2}$O$_{6}$OH)$_{2}$. Finally, Sec. \ref{conclusion} offers concluding remarks and summarizes the main outcomes of the study.

\section{Ising-Heisenberg trimer chain and its density operator}
\label{model}

\begin{figure}[t!]
\begin{center}
\includegraphics[width=0.45\textwidth]{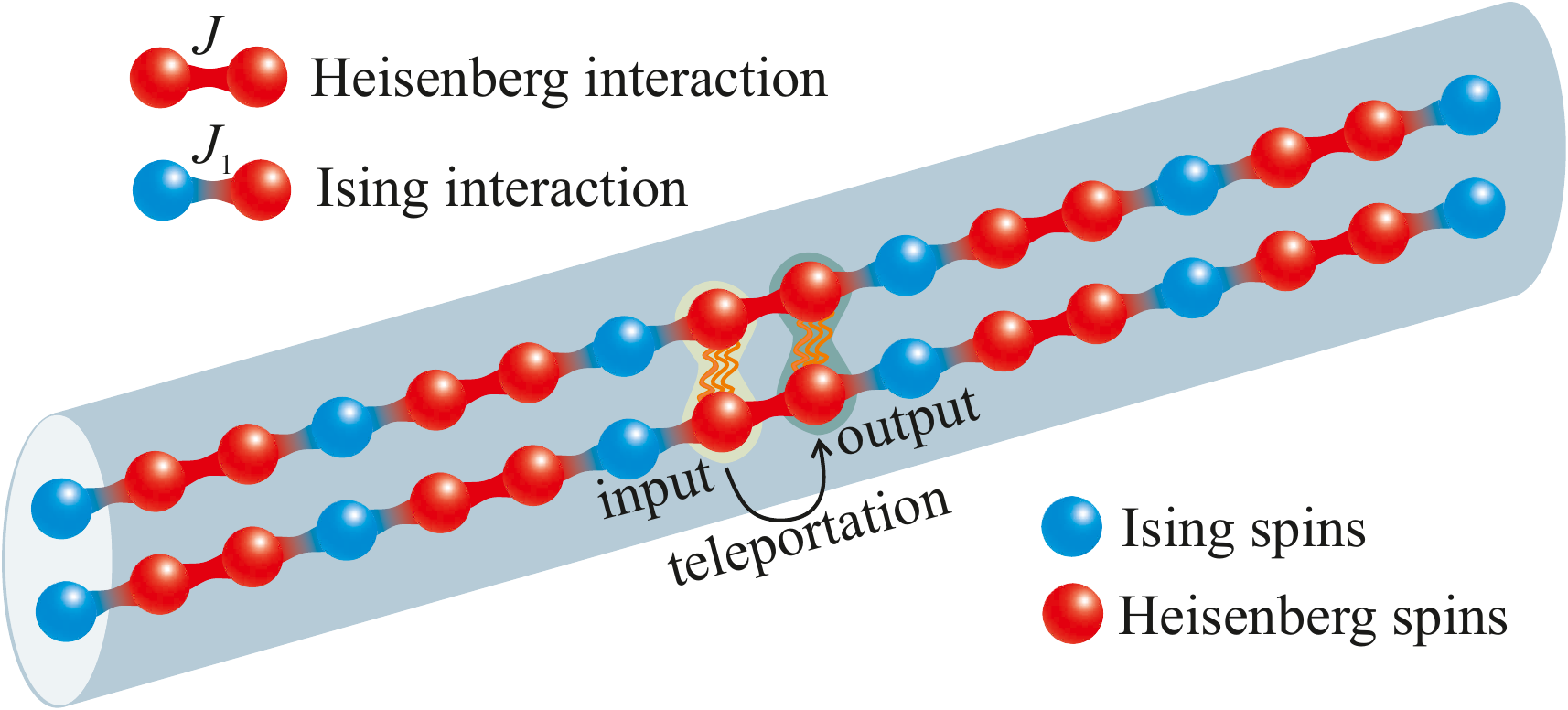}
\end{center}
\vspace{-0.7cm}
\caption{A schematic illustration of two independent spin-1/2 Ising-Heisenberg trimerized chains providing resource for the quantum communication channel. Blue spheres indicate lattice positions of the Ising spins, while red spheres represent lattice positions of the Heisenberg spins. The communication channel allows a quantum teleportation of an entangled two-qubit state prepared between two Heisenberg spins from two different dimers to their nearest-neighbour Heisenberg spins.}
\label{lattice}
\end{figure}

Consider a quantum channel formed by two independent spin-1/2 Ising-Heisenberg trimer chains schematically illustrated in Fig.~\ref{lattice}. The Hamiltonian describing a single spin-1/2 Ising-Heisenberg trimer chain is given by the following expression \cite{JO2003,JO2017}:
\begin{eqnarray}
\hat{\mathcal{H}} &=&J\sum_{k=1}^{N}\widehat{\mathbf{S}}_{2,k}\cdot \widehat{%
\mathbf{S}}_{3,k} + J_{1}\sum_{k=1}^{N}\left( \hat{s}^{z}_{1,k} \hat{S}^{z}_{2,k}
+\hat{S}_{3,k}^{z}\hat{s}^{z}_{1,k+1}\right)  \notag \\
&&-h\sum_{k=1}^{N}\left(\hat{s}^{z}_{1,k} + \hat{S}^{z}_{2,k}+\hat{S}^{z}_{3,k}\right),  \label{HA}
\end{eqnarray}%
where $\widehat{\mathbf{S}}_{2,k}\cdot \widehat{\mathbf{S}}_{3,k}=\hat{S}_{2,k}^{x}\hat{S}_{3,k}^{x}+\hat{S}_{2,k}^{y}\hat{S}_{3,k}^{y}+\hat{S}_{2,k}^{z}\hat{S}_{3,k}^{z}$ represents the scalar product of the standard spin-1/2 operators $\widehat{\mathbf{S}}_{\alpha,k} \equiv (\hat{S}^{x}_{\alpha,k}, \hat{S}^{y}_{\alpha,k}, \hat{S}^{z}_{\alpha,k})$ ($\alpha =2,3$) assigned to the quantum Heisenberg spins. The Hamiltonian (\ref{HA}) involves the spin operators $\hat{s}^{z}_{1,k}$ associated with the Ising spins having eigenvalues $s^{z}_{1,k}=\pm 1/2$. The coupling constant $J$ denotes the isotropic exchange interaction between the nearest-neighbor Heisenberg spins, whereas the coupling constant $J_{1}$ accounts for the anisotropic exchange interaction between the nearest-neighbor Ising and Heisenberg spins. Finally, the last term $h$ represents the standard Zeeman energy of the Ising and Heisenberg spins in the presence of an external magnetic field. For computational convenience, the total Hamiltonian (\ref{HA}) of the spin-1/2 Ising-Heisenberg trimerized chain can be reformulated as a sum of bond Hamiltonians:
\begin{equation}
\hat{\mathcal{H}}=\sum_{k=1}^{N}\hat{\mathcal{H}}_{k},
\end{equation}%
whereby each bond Hamiltonian $\hat{\mathcal{H}}_{k}$ can be decomposed into two parts:
\begin{equation}
\hat{\mathcal{H}}_{k}=-\frac{h}{2}(\hat{s}_{1,k}^{z}+\hat{s}_{1,k+1}^{z})+\hat{\mathcal{H}}_{k}^{\prime}.  
\label{HK1}
\end{equation}
The first term in the bond Hamiltonian (\ref{HK1}) describes the Zeeman energy of the two Ising spins within the $k$-th unit cell, whereas the second term $\hat{\mathcal{H}}_{k}^{\prime}$ includes all the interaction terms associated with the $k$-th pair of Heisenberg spins:
\begin{equation}
\hat{\mathcal{H}}_{k}^{\prime}=J \widehat{\mathbf{S}}_{2,k}\cdot \widehat{\mathbf{S}}_{3,k} 
-h_{2,k}\hat{S}_{2,k}^{z}-h_{3,k}\hat{S}_{3,k}^{z}.  \label{HK2}
\end{equation}%
Here, we have introduced the following notation for 'local' fields acting on the $k$-th Heisenberg spin pair:
\begin{equation}
h_{2,k}=h-J_{1}\hat{s}_{1,k}^{z} \text{ \ \ and \ \ }
h_{3,k}=h-J_{1}\hat{s}_{1,k+1}^{z}.
\end{equation}
The eigenvalues of the second part of the bond Hamiltonian (\ref{HK1}) can be obtained by a straightforward diagonalization of the Hamiltonian $\hat{\mathcal{H}}_{k}^{\prime}$ in a standard basis of the $k$-th Heisenberg dimer $\{\left\vert \uparrow \right\rangle_{2,k} \left\vert \uparrow \right\rangle_{3,k}, \left\vert \uparrow \right\rangle_{2,k} \left\vert \downarrow \right\rangle_{3,k}, \left\vert \downarrow \right\rangle_{2,k} \left\vert \uparrow \right\rangle_{3,k}, \left\vert \downarrow \right\rangle_{2,k} \left\vert \downarrow \right\rangle_{3,k}\}$:
\begin{eqnarray}
\varepsilon_{k,1}(s_{1,k},s_{1,k+1}) \!&=&\!
\frac{J}{4}+\frac{J_{1}}{2}(s_{1,k}^{z}+s_{1,k+1}^{z})-h,  \label{E1} \\
\varepsilon_{k,2}(s_{1,k},s_{1,k+1}) \!&=&\!
\frac{J}{4}-\frac{J_{1}}{2}(s_{1,k}^{z}+s_{1,k+1}^{z})+h,  \label{E2} \\
\varepsilon_{k,3}(s_{1,k},s_{1,k+1}) \!&=&\!-\frac{J}{4}
+\frac{1}{2}\sqrt{J_{1}^{2}(s_{1,k}^{z}\!-\!s_{1,k+1}^{z})^{2}\!+\!J^{2}}, \label{E3} \\
\varepsilon_{k,4}(s_{1,k},s_{1,k+1}) \!&=&\!-\frac{J}{4}
-\frac{1}{2}\sqrt{J_{1}^{2}(s_{1,k}^{z}\!-\!s_{1,k+1}^{z})^{2}\!+\!J^{2}}.
\label{E4}
\end{eqnarray}%
The eigenvectors of the $k$-th Heisenberg dimer, which correspond to the eigenvalues (\ref{E1})-(\ref{E4}), read as follows: 
\begin{eqnarray}
\left\vert \psi _{k,1}\right\rangle &=& \left\vert \uparrow \right\rangle_{2,k} \left\vert \uparrow \right\rangle_{3,k}, \label{P1} \\
\left\vert \psi _{k,2}\right\rangle &=& \left\vert \downarrow \right\rangle_{2,k} \left\vert \downarrow \right\rangle_{3,k},  \label{P2} \\
\left\vert \psi _{k,3}\right\rangle &=& a_{k}^{+} \left\vert \uparrow \right\rangle_{2,k} \left\vert \downarrow \right\rangle_{3,k} 
+ a_{k}^{-} \left\vert \downarrow \right\rangle_{2,k} \left\vert \uparrow \right\rangle_{3,k},  \label{P3} \\
\left\vert \psi _{k,4}\right\rangle &=& a_{k}^{-} \left\vert \uparrow \right\rangle_{2,k} \left\vert \downarrow \right\rangle_{3,k} 
- a_{k}^{+} \left\vert \downarrow \right\rangle_{2,k} \left\vert \uparrow \right\rangle_{3,k},  \label{P4}
\end{eqnarray}%
where the probability amplitudes $a_{k}^{\pm}$ entering into the last two eigenvectors (\ref{P3})-(\ref{P4}) are given by:
\begin{equation}
a_{k}^{\mp }=\frac{1}{\sqrt{2}}\sqrt{1\mp \frac{J_{1}({s}_{1,k}^{z}-{s}_{1,k+1}^{z})}
{\sqrt{J_{1}^{2}({s}_{1,k}^{z}-{s}_{1,k+1}^{z})^{2}+J^{2}}}}.
\end{equation}

The initial step in deriving the density operator of the spin-1/2 Ising-Heisenberg trimer chain is the calculation of the partition function, which can be efficiently carried out within the transfer-matrix formalism \cite{BAXTER1982}. Due to a validity of the commutation relation between bond Hamiltonians $[\hat{\mathcal{H}}_{k},\hat{\mathcal{H}}_{l}] = 0$, the partition function can be partially factorized into the following product:
\begin{equation}
\mathcal{Z}=\sum_{\{s_{1,k}^{z}\}}\prod\limits_{k=1}^{N}\mathrm{Tr}_{S_{2,k}}\mathrm{Tr}_{S_{3,k}}\exp \left( -\beta \hat{\mathcal{H}}_{k}\right),  
\label{Z}
\end{equation}
where $\beta =1/(k_{\rm B}T)$, $k_{\rm B}$ is the Boltzmann's constant, $T$ is the absolute temperature, the summation $\sum_{\{s_{1,k}^{z}\}}$ runs over all possible configurations of the Ising spins and the symbols $\mathrm{Tr}_{S_{2,k}}\mathrm{Tr}_{S_{3,k}}$ denote a trace over spin degrees of freedom of the $k$-th couple of Heisenberg spins. After performing the latter trace over degrees of freedom of the Heisenberg spin pair the partition function of the spin-1/2 Ising-Heisenberg trimer chain can be expressed in terms of the Boltzmann factor $\mathcal{T}({s}_{1,k}^{z},{s}_{1,k+1}^{z})$ depending on two nearest-neighbour Ising spins $s_{1,k}^{z}$ and $s_{1,k+1}^{z}$:
\begin{equation}
\mathcal{Z}=\sum_{\{s_{1,k}^{z}\}}\prod\limits_{k=1}^{N}\mathcal{T}\left({s}_{1,k}^{z},{s}_{1,k+1}^{z}\right).
\label{PF}
\end{equation}
The explicit form of the Boltzmann factor $\mathcal{T}({s}_{1,k}^{z},{s}_{1,k+1}^{z})$ directly follows from an energy spectrum (\ref{E1})-(\ref{E4}) of the $k$-th bond Hamiltonian:
\begin{eqnarray}
&& \mathcal{T}\left({s}_{1,k}^{z},{s}_{1,k+1}^{z}\right)=\mathrm{Tr}_{S_{2,k}}\mathrm{Tr}_{S_{3,k}}\exp \left( -\beta \hat{\mathcal{H}}_{k}\right) \nonumber \\
&& = \exp \left[\frac{\beta h}{2} \left({s}_{1,k}^{z} + {s}_{1,k+1}^{z}\right) \right] \sum_{l=1}^4 \exp(-\beta \varepsilon_{k,l}) \nonumber \\
&& = 2 {\rm e}^{\frac{\beta h}{2} ({s}_{1,k}^{z} + {s}_{1,k+1}^{z})} \! \left\{ {\rm e}^{-\frac{\beta J}{4}} \! \cosh \! \left[\frac{\beta J_1}{2} ({s}_{1,k}^{z}\!+\!{s}_{1,k+1}^{z}) - \beta h \right] \right. \nonumber \\
&& \left. \qquad \qquad + {\rm e}^{\frac{\beta J}{4}} \cosh \left[\frac{\beta}{2} \sqrt{J_1^2 ({s}_{1,k}^{z}\!-\!{s}_{1,k+1}^{z})^2 \!+\! J^2} \right]\right\}.
\label{TM}
\end{eqnarray}
The Boltzmann factor $\mathcal{T}({s}_{1,k}^{z},{s}_{1,k+1}^{z})$ can be eventually viewed as the transfer matrix and hence, one may perform in Eq. (\ref{PF}) a consecutive summation over spin states of the Ising spins to express the partition function in terms of the transfer-matrix eigenvalues:
\begin{equation}
\mathcal{Z}=\mathrm{Tr} \, \mathcal{T}^{N} = \lambda _{+}^{N}+\lambda _{-}^{N},
\label{EPF}
\end{equation}
which can be defined as follows:
\begin{equation}
\lambda_{\mp} = \frac{1}{2} \left[{T}_{1}+{T}_{2}\mp \sqrt{\left({T}_{1}-{T}_{2}\right) ^{2}+4{T}_{0}^2}\right].
\label{eig}
\end{equation}
The transfer-matrix eigenvalues (\ref{eig}) are expressed through three Boltzmann weights $T_1 = \mathcal{T}(+\frac{1}{2},+\frac{1}{2})$, $T_2 = \mathcal{T}(-\frac{1}{2},-\frac{1}{2})$, and $T_0=\mathcal{T}(+\frac{1}{2},-\frac{1}{2})=\mathcal{T}(-\frac{1}{2},+\frac{1}{2})$, which can be obtained from the transfer matrix (\ref{TM}) by considering all four available spin states of two Ising spins $s_{1,k}^{z}$ and $s_{1,k+1}^{z}$ involved therein: 
\begin{eqnarray}
{T}_1 &=& 2 {\rm e}^{\frac{\beta h}{2}} \! \left[{\rm e}^{-\frac{\beta J}{4}} \cosh \! \left(\frac{\beta J_1}{2} \!-\! \beta h \right) 
+ {\rm e}^{\frac{\beta J}{4}} \cosh \left(\frac{\beta J}{2} \right)\right], \nonumber \\
{T}_2 &=& 2 {\rm e}^{-\frac{\beta h}{2}} \! \left[{\rm e}^{-\frac{\beta J}{4}} \cosh \! \left(\frac{\beta J_1}{2} \!+\! \beta h \right) 
+ {\rm e}^{\frac{\beta J}{4}} \cosh \left(\frac{\beta J}{2} \right)\right], \nonumber \\
{T}_0 &=& 2 \! \left[{\rm e}^{-\frac{\beta J}{4}} \cosh \! \left(\beta h \right) 
+ {\rm e}^{\frac{\beta J}{4}} \cosh \left(\frac{\beta}{2} \sqrt{J_1^2 + J^2} \right)\right]\!\!. 
\label{TME}
\end{eqnarray}
In the thermodynamic limit $N\rightarrow \infty$, the partition function of the spin-1/2 Ising-Heisenberg trimer chain is determined solely by the dominant eigenvalue of the transfer-matrix $\mathcal{Z}=\lambda_{+}^{N}$. With this in mind, one can derive an exact expression for the free energy (per unit cell) of the spin-1/2 Ising-Heisenberg trimer chain:
\begin{eqnarray}
f &=& - k_{\rm B} T \ln \lambda _{+}  \label{GFE}  \\
  &=& k_{\rm B} T \ln 2 -  k_{\rm B} T \ln \left[T_1 + T_2 + \sqrt{(T_1-T_2)^2 + 4T_0^2} \right].
\nonumber
\end{eqnarray}
Similarly, one can easily derive exact expressions for the mean expectation values of the single-site magnetization of the Ising spins:
\begin{equation}
m_{\mathrm{I}} = \langle s_{1,k}^{z} \rangle = \frac{1}{2} \frac{T_1 - T_2}{\sqrt{(T_1-T_2)^2 + 4T_0^2}}
\label{MI}
\end{equation}
and the pair correlation function between the nearest-neighbour Ising spins:
\begin{equation}
\varepsilon_{\mathrm{I}} = \langle s_{1,k}^{z} s_{1,k+1}^{z} \rangle 
= m_{\mathrm{I}}^2 + \left( \frac{1}{4} - m_{\mathrm{I}}^2 \right) \frac{\lambda_{-}}{\lambda_{+}}.
\label{EI}
\end{equation}

Now, we may proceed to evaluation of the density operator. The overall density operator of the spin-1/2 Ising-Heisenberg trimerized chain is defined by the formula:
\begin{equation}
\hat{\varrho} = \frac{1}{\mathcal{Z}} \exp(-\beta \hat{\mathcal{H}}), 
\label{ODO}
\end{equation}
whereby the reduced density operator of the $k$-th Heisenberg spin pair $\hat{\varrho}^k$ can be obtained from this total density operator $\hat{\varrho}$ by tracing out spin degrees of freedom of all other spins:
\begin{eqnarray}
\hat{\varrho}^k = \frac{1}{\mathcal{Z}} \sum_{\{s_{1,j}^{z}\}} \mathrm{Tr}_{\{S_{2,j}\}_{j \neq k}}\mathrm{Tr}_{\{S_{3,k}\}_{j \neq k}} \exp(-\beta \hat{\mathcal{H}}).
\end{eqnarray}
After a straightforward algebraic manipulation one may derive the following formula for the reduced density operator of the $k$-th Heisenberg spin pair:
\begin{eqnarray}
\hat{\varrho}^k &=& \left\langle \frac{\exp(-\beta \hat{\mathcal{H}}_k)}{\mathrm{Tr}_{S_{2,k}}\mathrm{Tr}_{S_{3,k}} \exp(-\beta \hat{\mathcal{H}}_k)} \right\rangle \nonumber \\
&=& \left\langle \frac{\sum_{l=1}^4 \exp(-\beta \varepsilon_{k,l}) \left\vert \psi_{k,l}\right\rangle \left\langle \psi _{k,l}\right\vert}{\sum_{l=1}^4 \exp(-\beta \varepsilon_{k,l})} \right\rangle.
\end{eqnarray}
The reduced density matrix, which affords a matrix representation of the reduced density operator $\hat{\varrho}^k$ in a standard basis of the $k$-th Heisenberg spin pair, has the following structure:
\begin{equation}
\varrho^k =%
\begin{pmatrix}
\varrho_{11}^k & 0 & 0 & 0 \\
0 & \varrho_{22}^k & \varrho_{23}^k & 0 \\
0 & \varrho_{32}^k & \varrho_{33}^k & 0 \\
0 & 0 & 0 & \varrho_{44}^k
\end{pmatrix}.  
\label{RM}
\end{equation}
All non-vanishing elements of the reduced density matrix (\ref{RM}) can be defined through the unique formula:
\begin{eqnarray}
\varrho_{ij}^k &=& \frac{1}{4}\left[F_{ij}(1) + F_{ij}(-1) + 2 F_{ij}(0)\right] \nonumber \\
&+&\left[F_{ij}(1) + F_{ij}(-1) - 2 F_{ij}(0)\right] \varepsilon_{\mathrm{I}} \nonumber \\
&+& \left[F_{ij}(1) - F_{ij}(-1) \right] m_{\mathrm{I}},
\label{dme}
\end{eqnarray}
which depends on the single-site magnetization of the Ising spins given by Eq. (\ref{MI}), the pair correlation function between the nearest-neighbour Ising spins given by Eq. (\ref{EI}), as well as, the coefficients $F_{ij}(x)$ explicitly quoted in Appendix \ref{app}. In what follows, the derived density matrix (\ref{RM}) of the spin-1/2 Ising-Heisenberg trimerized chain will be used to investigate entanglement and teleportation through a  quantum channel composed from two spin-1/2 Heisenberg dimers.

\section{Quantum entanglement and teleportation}
\label{QTF}

During the teleportation process, the initial input state, applied to a pair of qubits belonging to two different Heisenberg dimers of the quantum channel, is destroyed. The output state, however, is reconstructed on the remaining pair of qubits from opposite ends of these two Heisenberg dimers through a local measurement process using linear operators. For this study, we consider the input state to be an arbitrary two-qubit pure state:

Now, let us turn our attention to the teleportation protocol designed for a quantum communication channel, which is formed by a pair of Heisenberg spin dimers selected from two distinct spin-1/2 Ising-Heisenberg trimer chains as schematically illustrated in Fig.~\ref{lattice}. It is assumed that these two spin-1/2 Ising-Heisenberg trimer chains are spatially well separated from each other to eliminate any potential coupling between them, which in turn means that two Heisenberg dimers creating this quantum channel are eventually independent of one another. The quantum teleportation of an entangled two-qubit pure state via the thermal mixed state of this quantum channel effectively modeled as a generalized depolarizing channel \cite{BOWEN2001,HORODECKI1999}. While the initial input state applied to a pair of qubits belonging to two distinct spin-1/2 Heisenberg dimers of the quantum channel is destroyed during the quantum teleportation, the output state is reconstructed on the remaining pair of qubits from opposite ends of these two spin-1/2 Heisenberg dimers through a local measurement using linear operators. Specifically, we consider the input state to be an arbitrary two-qubit pure state:
\begin{equation}
\left\vert \psi _{\mathrm{in}}\right\rangle =\cos \left( \frac{\theta }{2}\right) \left\vert 1\right\rangle \left\vert 0\right\rangle +\exp \left( i\varphi \right) \sin \left(\frac{\theta }{2}\right) \left\vert 0\right\rangle \left\vert 1\right\rangle.
\end{equation}
The input state is expressed in the natural basis of qubits $\{ \vert 0\rangle, \vert 1\rangle \}$, which should not be confused with the standard basis previously used for the spin-1/2 Heisenberg dimers. The parameter $\theta \left( 0\leq \theta \leq \pi \right)$ serves as a mixing angle that determines the probability amplitudes within the quantum superposition of two basis states $\vert 1\rangle \vert 0\rangle$ and $\vert 0\rangle \vert 1\rangle$, while the parameter $\varphi \left( 0\leq \varphi \leq 2\pi \right)$ represents the relative phase of this pure state. The concurrence of the input state $\mathcal{C}_{\mathrm{in}}$, which quantifies the degree of entanglement of the initial state to be teleported, can be straightforwardly calculated from the density operator of the input state $\hat{\varrho}_{\mathrm{in}}=\left\vert \psi _{\mathrm{in}}\right\rangle \left\langle \psi _{\mathrm{in}}\right\vert$ following the procedure established by Hill and Wootters \cite{HILL1997,WOOTTERS1998}:
\begin{equation}
\mathcal{C}_{\mathrm{in}}=\left\vert \sin \left( \theta \right) \right\vert.
\end{equation}%
On the other hand, the concurrence of the quantum channel $\mathcal{C}_{\mathrm{ch}}$ affording the quantum resource for the teleportation of an entangled two-qubit state through two Heisenberg dimers of the spin-1/2 Ising-Heisenberg trimer chains can be directly determined from the elements of the density matrix of this quantum channel $\varrho^k$ given by Eq.~(\ref{dme}):
\begin{equation}
\mathcal{C}_{\mathrm{ch}}=2 {\rm max} \left\{ 0, \left\vert \varrho_{23}^k \right\vert - \sqrt{\varrho_{11}^k \varrho_{44}^k}\right\}.
\end{equation}
When the two-qubit pure input state $\hat{\varrho}_{\mathrm{in}}=\left\vert \psi _{\mathrm{in}}\right\rangle \left\langle \psi_{\mathrm{in}}\right\vert$ is transmitted via the mixed thermal state of the quantum channel formed by two spin-1/2 Heisenberg dimers belonging to two distinct spin-$1/2$ Ising-Heisenberg trimer chains, the teleportation protocol comprising joint measurements and local unitary transformations applied to the input state $\hat{\varrho}_{\mathrm{in}}$ the following replica output state:
\begin{equation}
\hat{\varrho}_{\mathrm{out}}=\sum_{i=0}^3 \sum_{i=0}^3 p_i p_j\left(\hat{\sigma}_{i}\otimes \hat{\sigma}_{j}\right) \hat{\varrho}_{\mathrm{in}}\left(\hat{\sigma}_{i}\otimes \hat{\sigma}_{j}\right),
\end{equation}%
where $\hat{\sigma}_{0}=\hat{I}$, $\hat{\sigma}_{1}=\hat{\sigma}_{x}$, $\hat{\sigma}_{2}=\hat{\sigma}_{y}$, and $\hat{\sigma}_{3}=\hat{\sigma}_{z}$ are the identity operator and three spatial components of the Pauli spin operator, respectively. The coefficients $p_{i}=\mathrm{Tr} \, [\hat{\varrho}^k \hat{P}_{i}]$ determine probability of the quantum channel to be in one out of the four fully entangled Bell states $\left\vert \Psi^{\pm}\right\rangle = \frac{1}{\sqrt{2}} \left(\left\vert 0\right\rangle \left\vert 1\right\rangle \pm \left\vert 1\right\rangle \left\vert 0\right\rangle \right)$ and $\left\vert \Phi^{\pm}\right\rangle = \frac{1}{\sqrt{2}} \left(\left\vert 0\right\rangle \left\vert 0\right\rangle \pm \left\vert 1\right\rangle \left\vert 1\right\rangle \right)$, whereby $\hat{P}_{i}$ denote the respective projection operators on the relevant Bell states $\hat{P}_{0}= \left\vert \Psi^{-}\right\rangle \left\langle \Psi^{-}\right\vert$, $\hat{P}_{1} = \left\vert \Phi^{-}\right\rangle \left\langle \Phi^{-}\right\vert$, $\hat{P}_{2} = \left\vert \Phi^{+}\right\rangle \left\langle \Phi ^{+}\right\vert$, $\hat{P}_{3}=\left\vert \Psi^{+}\right\rangle \left\langle \Psi^{+}\right\vert$. The density matrix of the output state $\rho_{\mathrm{out}}$ formally has the same matrix representation as the density matrix of the quantum channel $\varrho^k$ given by Eq. (\ref{RM}):
\begin{equation}
\varrho_{\mathrm{out}}=
\begin{pmatrix}
\varrho_{11}^{out} & 0 & 0 & 0 \\
0 & \varrho_{22}^{out} & \varrho_{23}^{out} & 0 \\
0 & \varrho_{32}^{out} & \varrho_{33}^{out} & 0 \\
0 & 0 & 0 & \varrho_{44}^{out}
\end{pmatrix},
\end{equation}%
whereby its non-zero matrix elements can be connected to that ones of the quantum channel through the following set of equations:
\begin{eqnarray}
\varrho_{11}^{out} &=&\varrho_{44}^{out}=2\varrho_{22}^k \left(\varrho_{11}^k+\varrho_{44}^k\right), \nonumber \\
\varrho_{22}^{out} &=& 4\cos^{2}\left(\frac{\theta}{2}\right) \left(\varrho_{22}^k \right)^{2}+\sin ^{2}\left(\frac{\theta}{2}\right) \left(\varrho_{11}^k+\varrho_{44}^k\right)^{2}\!\!, \nonumber \\
\varrho_{33}^{out} &=&4\sin^{2}\left(\frac{\theta}{2}\right) \left(\varrho_{22}^k\right)^{2}+\cos ^{2}\left(\frac{\theta}{2}\right) \left(\varrho_{11}^k+\varrho_{44}^k\right)^{2}\!\!, \nonumber \\
\varrho_{23}^{out} &=& 2\sin \left(\theta \right) \exp \left( -{\rm i} \varphi \right) \left(\varrho_{23}^k\right)^{2}\!\!, \quad \varrho_{32}^{out} = \left(\varrho_{23}^{out}\right)^{*}\!\!.
\end{eqnarray}%
The concurrence $\mathcal{C}_{\mathrm{out}}$ measuring a degree of the entanglement of the output state is determined as follows:
\begin{eqnarray}
\mathcal{C}_{\mathrm{out}} &=& 2 {\rm max} \left\{ 0, \left\vert \varrho_{23}^{out} \right\vert 
- \sqrt{\varrho_{11}^{out} \varrho_{44}^{out}}\right\} \nonumber \\
&=& 4 \max \left\{ 0,\left(\varrho_{23}^{k}\right)^{2}\mathcal{C}_{\mathrm{in}}
-\varrho_{22}^{k} \left(\varrho_{11}^{k}+\varrho_{44}^{k}\right) \right\}.
\end{eqnarray}
To assess the quality of this quantum teleportation protocol, one may calculate for the input and output states 
$\varrho_{\mathrm{in}}$ and $\varrho_{\mathrm{out}}$ the quantity fidelity defined by \cite{JOZSA1994,BOWDREY2002}:
\begin{equation}
\mathcal{F}=\left( \mathrm{Tr}\sqrt{\sqrt{\varrho_{\mathrm{in}}}\varrho_{\mathrm{out}}\sqrt{\varrho_{\mathrm{in}}}}\right)^{2}.
\end{equation}
If the input state to be teleported is a pure quantum state represented by the state vector $\left\vert \psi _{\mathrm{in}}\right\rangle$, the fidelity of the quantum teleportation can be determined as the quantum-mechanical expectation value of the thermal density operator for the output state: $\mathcal{F}= \left\langle \psi _{\mathrm{in}}\right\vert \hat{\varrho}_{\mathrm{out}} \left\vert \psi _{\mathrm{in}}\right\rangle$. After performing a series of algebraic manipulations, one acquires the following exact result for the fidelity of the quantum teleportation through the quantum channel composed from two spin-1/2 Heisenberg dimers embedded within the two spin-1/2 Ising-Heisenberg trimer chains:
\begin{equation}
\mathcal{F}=\frac{\sin^{2}\theta}{2} \! \left[\left(\varrho_{11}^{k} \!+\! \varrho_{44}^{k}\right)^{2}
 \!+\! 4 (\varrho_{23}^{k})^{2} \!-\! 4 (\varrho_{22}^{k})^{2}\right] \!+\! 4 (\varrho_{22}^{k})^{2}\!.
\end{equation}
The efficiency of quantum teleportation through this communication channel can also be quantified by the average fidelity 
$\mathcal{F}_{av}$ obtained by averaging the fidelity over all possible two-qubit quantum states \cite{JOZSA1994,BOWDREY2002}:
\begin{equation}
\mathcal{F}_{av} = \frac{1}{4\pi} \int_{0}^{2\pi} {\rm d} \varphi \int_{0}^{\pi} 
\mathcal{F} \sin \theta {\rm d} \theta.
\end{equation}%
The average fidelity $\mathcal{F}_{av}$ provides a comprehensive evaluation of the performance of teleportation protocol irrespective of the specific input state and can be expressed through the following formula obtained after performing the relevant integration:
\begin{equation}
\mathcal{F}_{av}=\frac{1}{3}\left[ \left(\varrho_{11}^{k}+\varrho_{44}^{k}\right)^{2}
+ 4 (\varrho_{23}^{k})^{2} - 4 (\varrho_{22}^{k})^{2} \right] + 4 (\varrho_{22}^{k})^{2}.
\end{equation}
To surpass performance of classical communication protocols, the average fidelity should exceed the threshold value of
$\mathcal{F}_{A} = 2/3$, which represents the upper bound for fidelity achievable within the classical regime 
\cite{JOZSA1994,BOWDREY2002}.

\section{Results and discussion}
\label{Results}

In this section, we provide a comprehensive analysis of quantum entanglement and quantum teleportation realized within the Heisenberg spin pairs of the spin-1/2 Ising-Heisenberg trimer chains. This analysis will be limited to a particular case of the spin-1/2 Ising-Heisenberg trimer chain with antiferromagnetic coupling constants $J>0$ and $J_{1}>0$, which are anticipated to showcase the most pronounced quantum features.

\begin{figure}[t]
\includegraphics[width=0.5\textwidth]{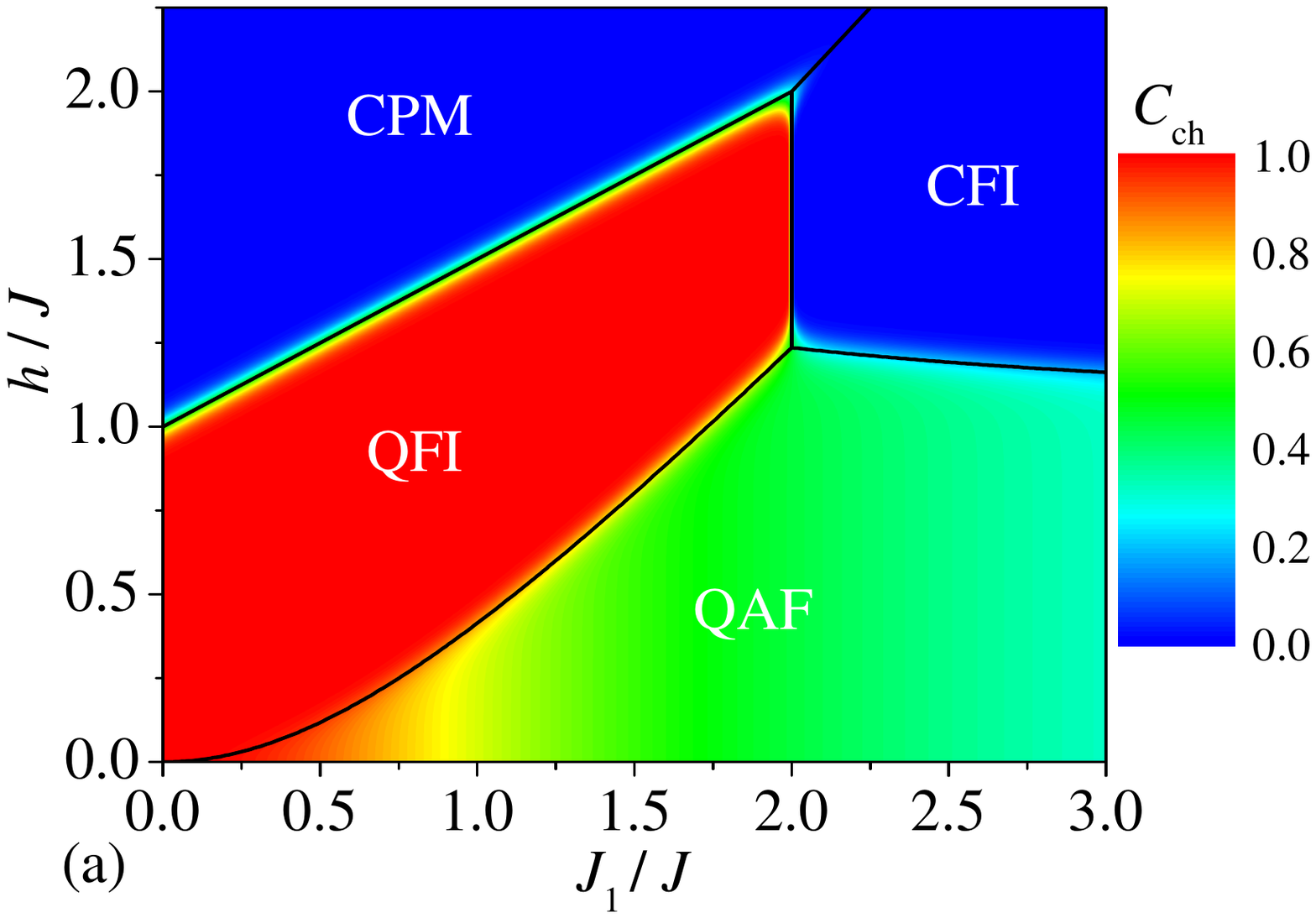}
\includegraphics[width=0.5\textwidth]{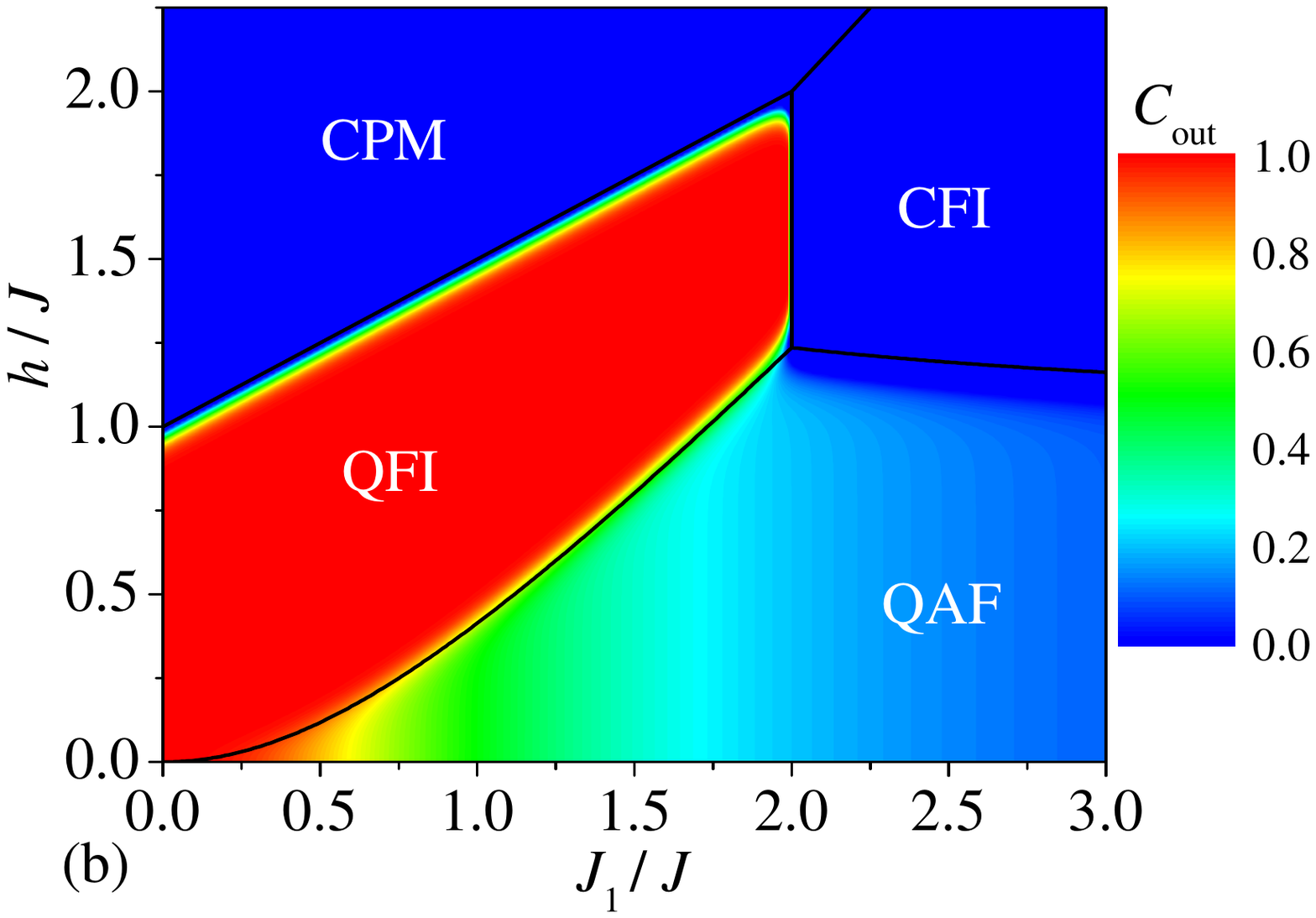}
\includegraphics[width=0.5\textwidth]{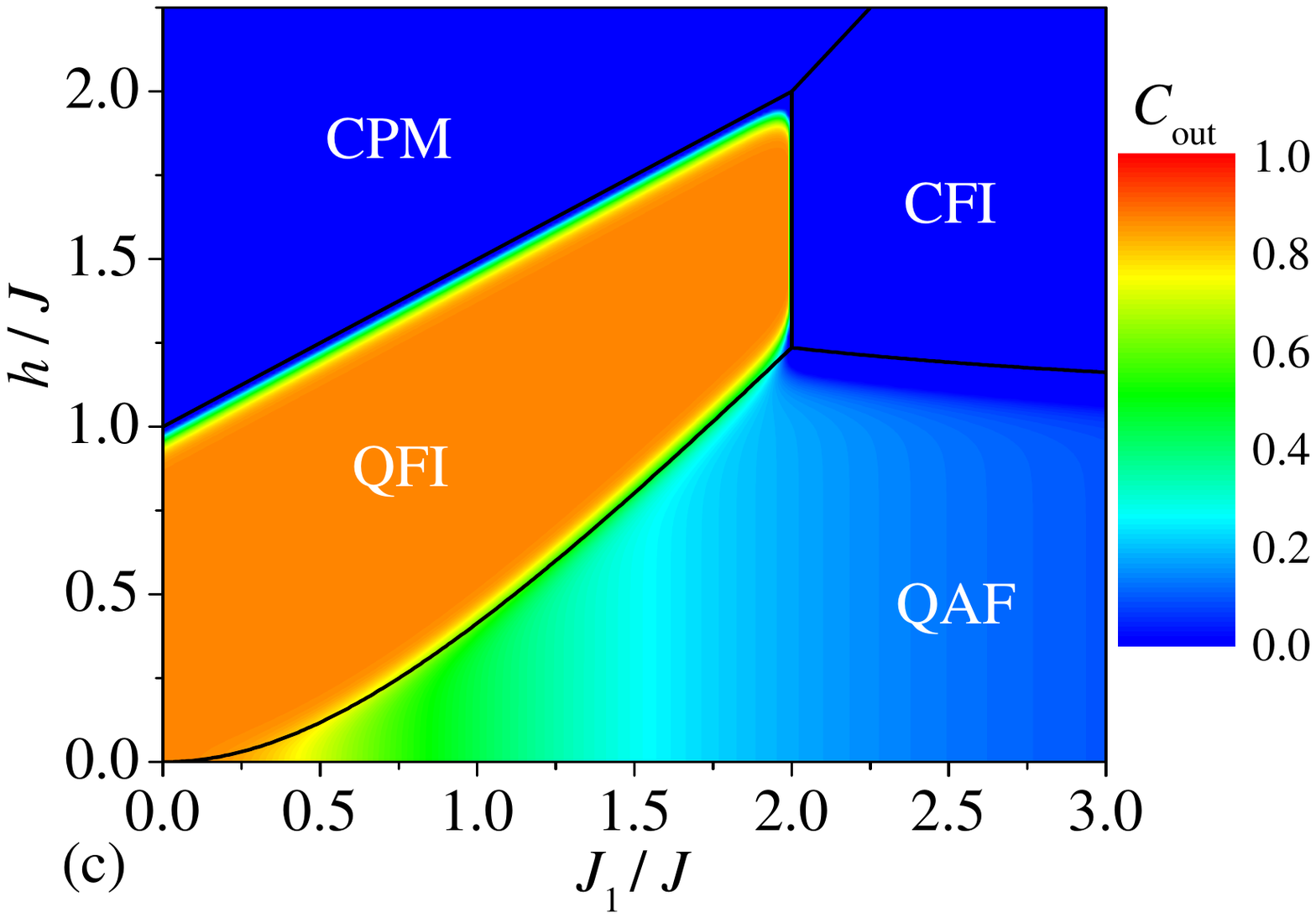}
\vspace{-0.4cm}
\caption{(a) A density plot of the concurrence of the quantum channel $\mathcal{C}_{ch}$ composed from two Heisenberg dimers of the spin-1/2 Ising-Heisenberg trimerized chains in the plane $J_{1}/J-h/J$ at sufficiently low temperature $k_{\rm B}T/J= 0.01$; (b)-(c) Density plots of the concurrence of the output state $\mathcal{C}_{out}$ in the plane $J_{1}/J-h/J$ obtained at low enough temperature $k_{\rm B}T/J= 0.01$ by implementing the teleportation protocol for two selected input states with the mixing angle: (b) $\protect\theta=\protect\pi/2$, (c) $\protect\theta=\protect\pi/3$. Black lines determine zero-temperature phase boundaries between different ground states (see the text for the notation).}
\label{cdenjh}
\end{figure}

To begin, we will examine in detail the bipartite entanglement emerging within the Heisenberg spin pairs of the spin-1/2 Ising-Heisenberg trimer chain, because the quantum entanglement represents a necessary prerequisite for the implementation of the quantum teleportation protocol. To this end, the concurrence of the Heisenberg dimers $\mathcal{C}_{ch}$, serving as a platform of the quantum communication channel, is plotted in Fig.~\ref{cdenjh}(a) within the parameter plane $J_{1}/J-h/J$. Notably, the concurrence $\mathcal{C}_{ch}$ undergoes abrupt changes across the phase boundaries separating four distinct ground states of the spin-1/2 Ising-Heisenberg trimer chain referred to as the quantum antiferromagnetic phase (QAF), the quantum ferrimagnetic phase (QFI), the classical ferrimagnetic phase (CFI), and the classical paramagnetic phase (CPM):  
\begin{eqnarray}
|\mbox{QAF}\rangle &\!\!=\!\!& \prod_{k = 1}^{N/2} \! |{\uparrow}\rangle_{1,2k\!-\!1}  
\! \left(c_{-} |{\uparrow}\rangle_{2,2k\!-\!1} |{\downarrow}\rangle_{3,2k\!-\!1} 
    \!\!-\!\! c_{+} |{\downarrow}\rangle_{2,2k\!-\!1}|{\uparrow}\rangle_{3,2k\!-\!1} \! \right) \nonumber \\
&& \quad \,\,\,\, |{\downarrow}\rangle_{1,2k}   
\left(c_{+} |{\uparrow}\rangle_{2,2k}|{\downarrow}\rangle_{3,2k} 
    \!-\! c_{-} |{\downarrow}\rangle_{2,2k}|{\uparrow}\rangle_{3,2k}\right), \label{QAF} \\
|\mbox{QFI}\rangle &\!\!=\!\!& \prod_{k = 1}^{N} |{\uparrow} \rangle_{1,k} 
\frac{1}{\sqrt{2}} \left(|{\uparrow}\rangle_{2,k}|{\downarrow}\rangle_{3,k} 
                       - |{\downarrow}\rangle_{2,k}|{\uparrow}\rangle_{3,k}\right), \label{QFI} \\
|\mbox{CFI}\rangle &\!\!=\!\!& \prod_{k = 1}^{N} |{\downarrow}\rangle_{1,k}  
|{\uparrow}\rangle_{2,k}|{\uparrow}\rangle_{3,k}, \, \, \label{CFI} \\
|\mbox{CPM}\rangle &\!\!=\!\!& \prod_{k = 1}^{N} |{\uparrow}\rangle_{1,k}  
|{\uparrow}\rangle_{2,k}|{\uparrow}\rangle_{3,k}. \label{CPM}
\end{eqnarray}
The probability amplitudes entering into the ground-state eigenvector $|\mbox{QAF}\rangle$ are given by: 
\begin{equation}
c_{\pm} = \frac{1}{\sqrt{2}}\sqrt{1 \pm \frac{J_{1}}{\sqrt{J_{1}^{2} + J^{2}}}}.
\end{equation}
The concurrence of the Heisenberg spin pairs is zero in the two factorizable classical ground states CFI and CPM described by the eigenvectors (\ref{CFI}) and  (\ref{CPM}), whereas the concurrence of the Heisenberg spin pairs becomes nonzero $\mathcal{C}_{ch}^{QAF} = \frac{J}{\sqrt{J_1^2 + J^2}}$ and $\mathcal{C}_{ch}^{QFI} = 1$ in the two quantum ground states QAF and QFI determined by the eigenvectors (\ref{QAF}) and (\ref{QFI}), respectively. The QFI ground state features a singlet-dimer state of the Heisenberg spin pairs characterized by a perfect bipartite entanglement, while the QAF ground state features only a partial bipartite entanglement of the Heisenberg spin pairs progressively weakening within singlet-dimer-like state upon increasing of the interaction ratio $J_1/J$. As the magnetic field favors the QFI ground state over the QAF ground state, one may eventually strengthen the bipartite entanglement within the quantum communication channel by optimal tuning of the magnetic field. 

Furthermore, our attention is focused on a detailed analysis of the concurrence of the output state $\mathcal{C}_{out}$, which quantifies the strength of the bipartite entanglement within the transmitted two-qubit state generated through the quantum communication channel formed by the two Heisenberg dimers of the spin-1/2 Ising-Heisenberg trimer chains. The output concurrence $\mathcal{C}_{out}$ is plotted in Fig.~\ref{cdenjh}(b) and (c) in the $J_{1}/J-h/J$ plane for two specific input states characterized by the mixing angles $\protect\theta=\protect\pi/2$ and $\protect\pi/3$ displaying a perfect and partial quantum entanglement, respectively. It is clear that the output concurrence $\mathcal{C}_{out}$ exhibits the same general trends as the concurrence of the quantum channel $\mathcal{C}_{ch}$. If the interaction ratio and magnetic field drive the quantum channel towards the QAF ground state, the output concurrence $\mathcal{C}_{out}$ is somewhat reduced  compared to the concurrence of the quantum channel $\mathcal{C}_{ch}$. However, the magnetic-field-driven strengthening of the output concurrence $\mathcal{C}_{out}$ can be still detected whenever the quantum channel is driven towards the QFI ground state.

\begin{figure}[t]
\includegraphics[width=0.5\textwidth]{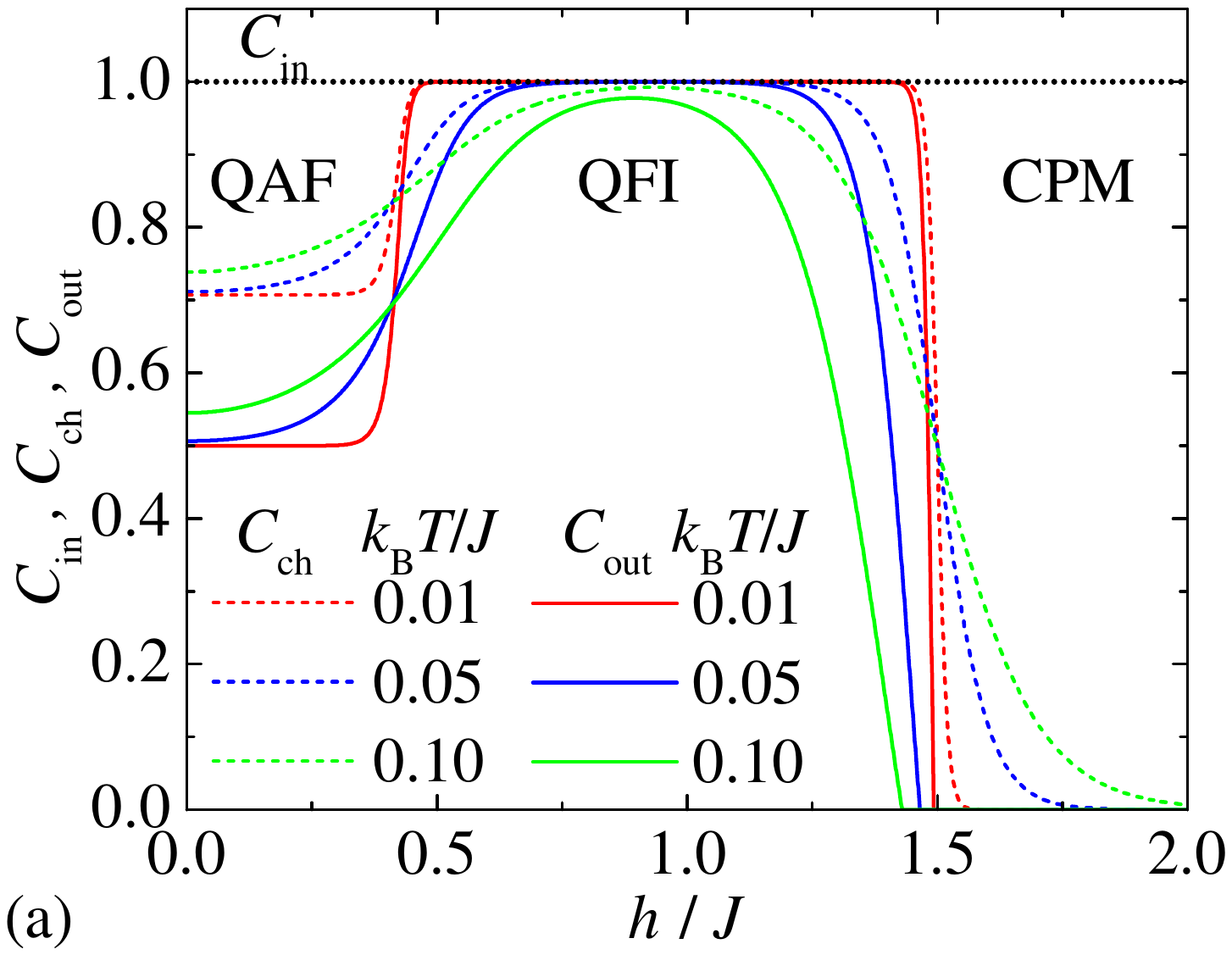}
\includegraphics[width=0.5\textwidth]{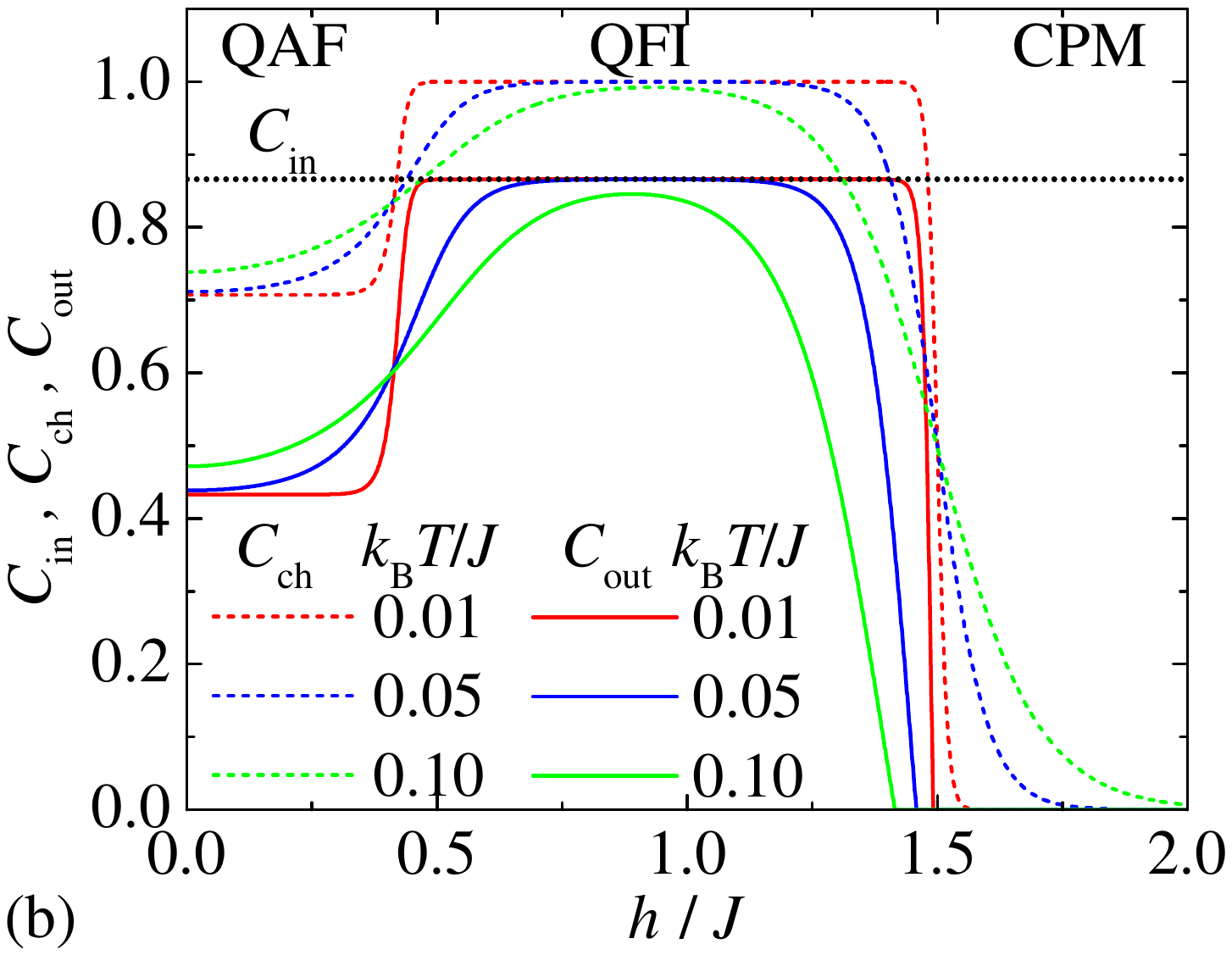}
\vspace{-0.4cm}
\caption{A comparison between the concurrence of the quantum channel $\mathcal{C}_{ch}$ and the concurrence of the output state $\mathcal{C}_{out}$ as a function of the magnetic field for one specific value of the interaction ratio $J_{1}/J=1$ and three different temperatures by assuming the input state with the mixing angle: (a) $\protect\theta=\protect\pi/2$, (b) $\protect\theta=\protect\pi/3$. A dotted line represents the concurrence $\mathcal{C}_{in}$ of the input state transmitted through the communication channel.}
\label{c2d}
\end{figure}

Next, let us explore how the transient enhancement of bipartite entanglement driven by the magnetic-field-induced transition from the QAF to the QFI ground state manifests itself at finite (non-zero) temperatures. For this purpose, we have illustrated in Fig.~\ref{c2d} typical magnetic-field dependencies of the concurrence of the quantum channel $\mathcal{C}_{ch}$ and the  concurrence of the output state $\mathcal{C}_{out}$ for a fixed value of the interaction ratio $J_{1}/J=1$ and a few selected temperatures. In agreement with the previous analysis, three distinct regimes of the concurrence are expected for the parameter regions corresponding to the three ground states QAF, QFI, and CPM. In the low-field regime $h/J\lesssim 0.4$ inherent to the QAF ground state, the concurrence of the quantum channel reaches a value of $\mathcal{C}_{ch}\approx 0.7$, while the concurrence of the output state $\mathcal{C}_{out}$ reaches only about half of the concurrence of the input state $\mathcal{C}_{in}$. In the range of moderate magnetic fields $0.4\lesssim h/J \lesssim 1.5$ pertinent to the QFI ground state, the concurrence of the quantum channel approaches at low enough temperatures the highest possible value $\mathcal{C}_{ch}=1$, which consequently facilitates reliable teleportation as reflected by coincidence of the concurrence of the input and output states found regardless of whether the transmitted quantum state is fully or only partially entangled. In the high-field regime $h/J \gtrsim 1.5$ relevant to the CPM ground state, the concurrence of the quantum channel tends to zero $\mathcal{C}_{ch}\approx 0$ at sufficiently low temperatures effectively preventing efficient quantum communication through this quantum channel though a small degree of thermal entanglement can be induced as the temperature increases. It is also evident from Fig.~\ref{c2d} that increasing temperature gradually reduces and smears out the sharp stepwise dependencies of both the concurrence of the quantum channel $\mathcal{C}_{ch}$ and the output state $\mathcal{C}_{out}$. 

\begin{figure}[t]
\includegraphics[width=0.5\textwidth]{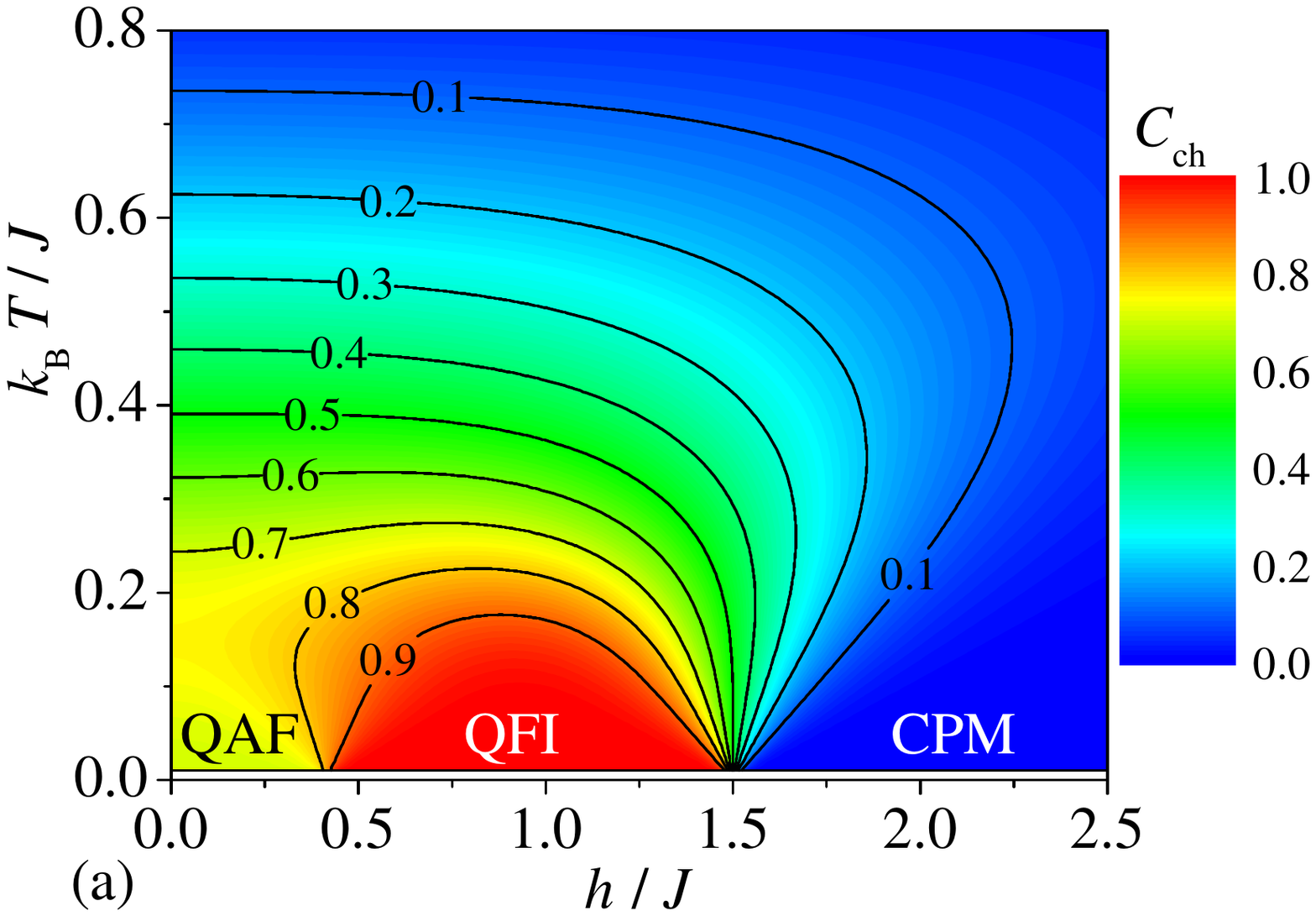}
\includegraphics[width=0.5\textwidth]{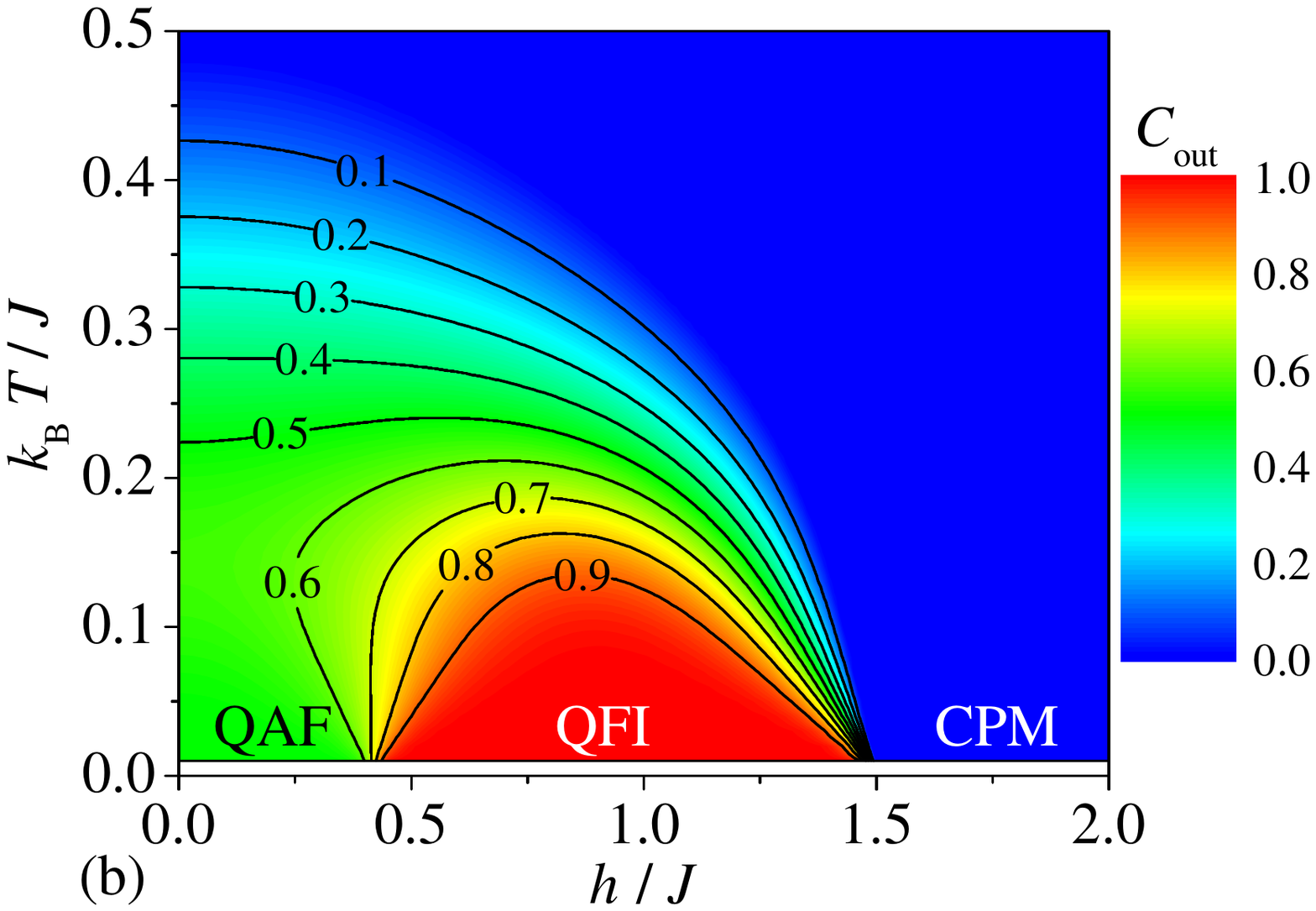}
\includegraphics[width=0.5\textwidth]{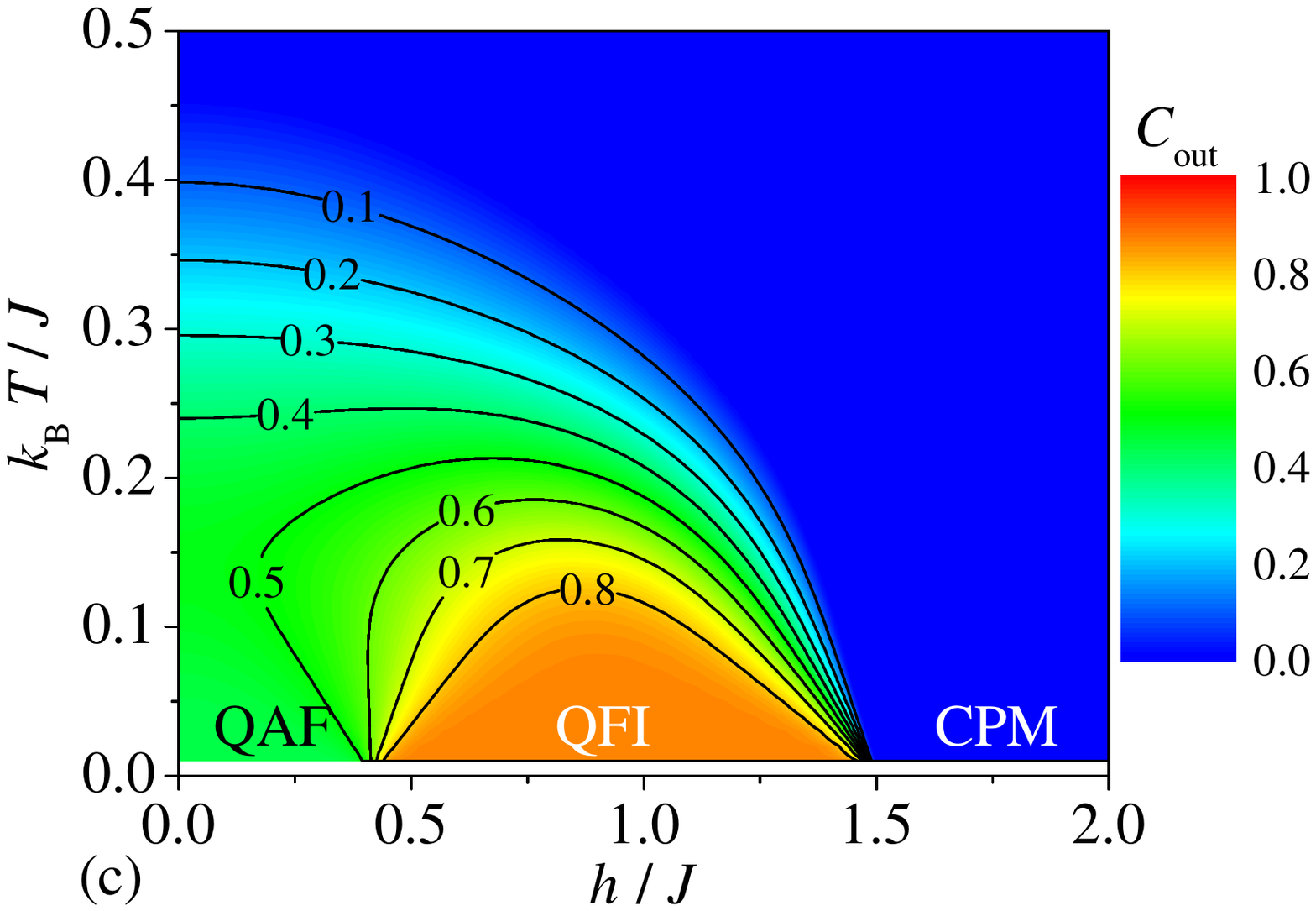}
\vspace{-0.4cm}
\caption{(a) A density plot of the concurrence of the quantum channel $\mathcal{C}_{ch}$ composed from two Heisenberg dimers of the spin-1/2 Ising-Heisenberg trimerized chains in the plane $h/J-k_{\rm B}T/J$ for the interaction ratio $J_{1}/J=1$; (b)-(c) Density plots of the concurrence of the output state $\mathcal{C}_{out}$ in the plane $h/J-k_{\rm B}T/J$ obtained for the interaction ratio $J_{1}/J=1$ by implementing the teleportation protocol for two different input states with the mixing angle: (b) $\protect\theta=\protect\pi/2$, (c) $\protect\theta=\protect\pi/3$.}
\label{c3d}
\end{figure}  

To bring an overall insight into the strength of thermal entanglement at finite temperatures, the concurrence of the quantum channel $\mathcal{C}_{ch}$ formed by two Heisenberg dimers of the spin-1/2 Ising-Heisenberg trimer chains and the concurrence of the output state are plotted in Fig.~\ref{c3d} in the magnetic field versus temperature plane. The relevant density plots clearly distinguish the QAF, QFI and CPM parameter regions due to sizable changes in the concurrence and thus provide an additional independent confirmation of the magnetic-field-driven enhancement of thermal entanglement observable at low enough temperatures $k_{\rm B}T/J \lesssim 0.2$. Although one detects in Fig.~\ref{c3d} the similar general trends, the concurrence of the quantum channel $\mathcal{C}_{ch}$ turns out to be much more resilient to increasing temperature compared to the concurrence of the output state $\mathcal{C}_{out}$. Indeed, the contour lines representing the same magnitude of concurrence can be found at nearly twice as high temperature for the quantum channel as for the output state. Furthermore, the threshold (sudden-death) temperature above which thermal entanglement completely vanishes is also notably higher for the quantum channel than for the output state. Another fundamental difference between $\mathcal{C}_{ch}$ and $\mathcal{C}_{out}$ lies in the remarkable behavior of the concurrence of the quantum channel, which displays in the high-field region $h/J \gtrsim 1.5$ a striking sudden-birth phenomenon when a relatively weak thermal entanglement is induced above the classical ground state CPM due to thermal activation of quantum excited states. Contrary to this, the thermally-assisted increase in the bipartite entanglement does not occur in the relevant density plots of the output concurrence as depicted in Fig.~\ref{c3d}(b) and (c) for two transmitted input states with the mixing angles $\theta =\pi/2$ and $\theta =\pi/3$, respectively.
 
\begin{figure}[t]
\includegraphics[width=0.5\textwidth]{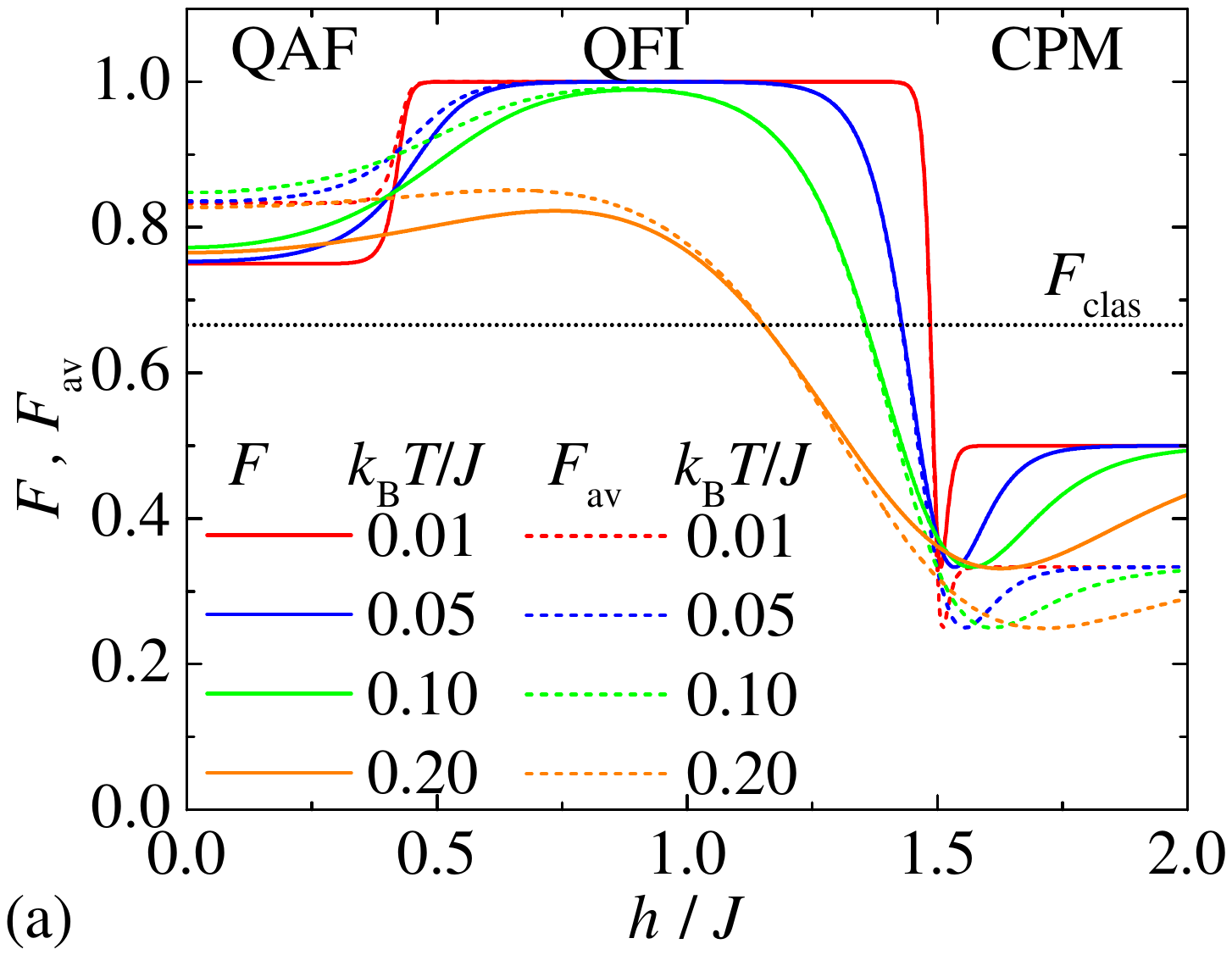} 
\includegraphics[width=0.5\textwidth]{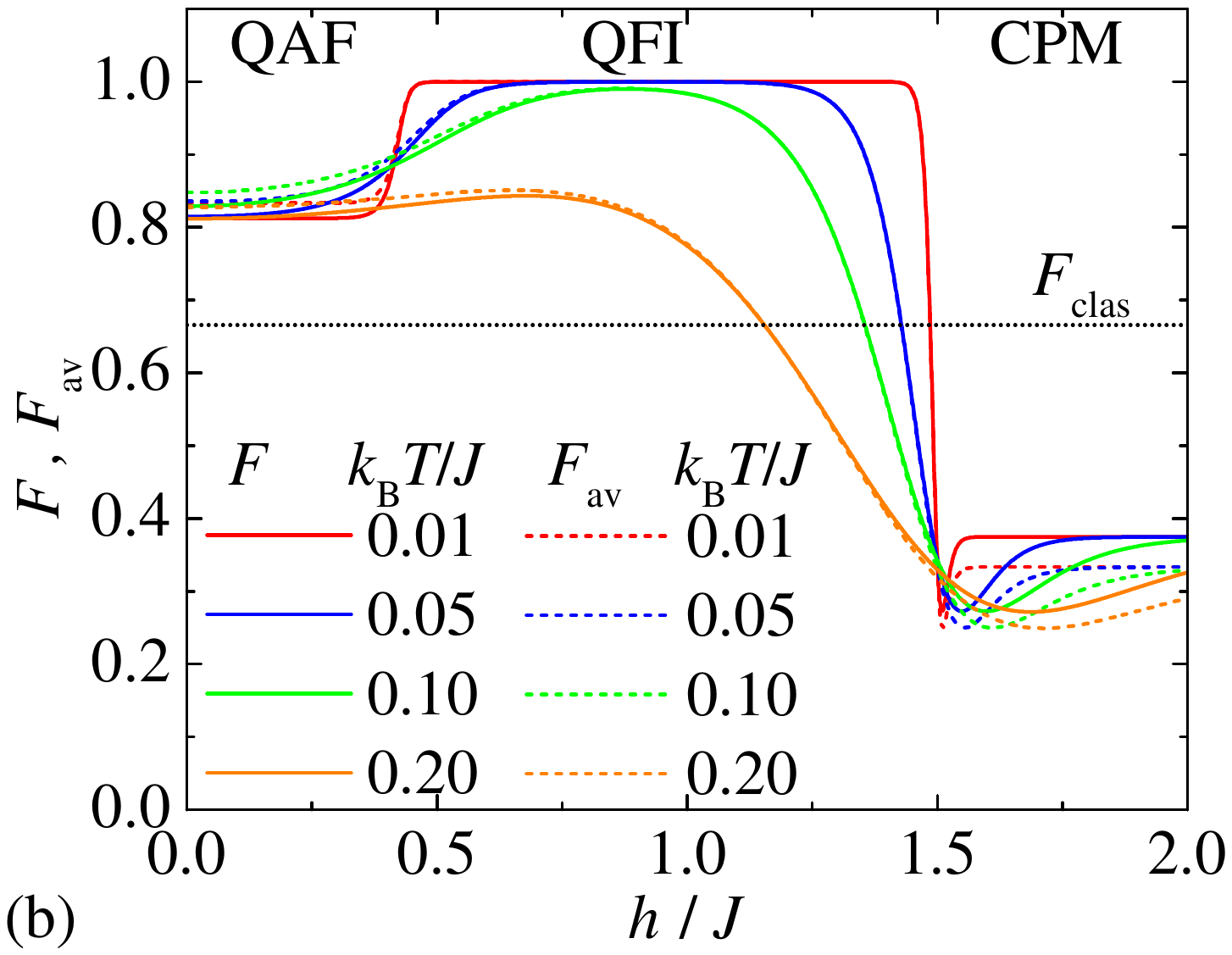} 
\vspace{-0.4cm}
\caption{Magnetic-field dependencies of the fidelity $\mathcal{F}$ (solid lines) and the average fidelity $\mathcal{F}_{av}$ (broken lines) for the fixed value of the interaction ratio $J_{1}/J=1$ and four selected temperatures. The panel (a) illustrates the fidelity for the teleportation of a fully entangled input state with the mixing angle $\protect\theta=\protect\pi/2$, while the panel (b) illustrates the fidelity for the teleportation of a partially entangled input state with the mixing angle $\protect\theta=\protect\pi/3$. A thin dotted line represents the threshold value $\mathcal{F}_{clas} = 2/3$ for a classical communication channel.}
\label{figf}
\end{figure}

To assess the reliability of quantum communication through the quantum channel formed by two Heisenberg dimers of the spin-1/2 Ising-Heisenberg trimer chains, the quantity fidelity is plotted in Fig.~\ref{figf} as a function of the magnetic field for one selected value of the interaction ratio $J_{1}/J = 1$ and four different temperatures. For comparison, Fig.~\ref{figf}(a) displays the fidelity for the quantum teleportation of a fully entangled two-qubit state with the mixing angle $\protect\theta=\protect\pi/2$ and the input concurrence $\mathcal{C}_{in} = 1$, while Fig.~\ref{figf}(b) depicts the fidelity for the quantum teleportation of a partially entangled two-qubit state with the mixing angle $\protect\theta=\protect\pi/3$ and the input concurrence $\mathcal{C}_{in} \approx 0.87$. In both cases one observes abrupt changes in fidelity at sufficiently low temperatures occurring around two magnetic fields $h/J \approx 0.4$ and $1.5$, which correspond at absolute zero temperature to magnetic-field-driven phase transitions between the QAF-QFI and QFI-CPM ground states, respectively. In the low-field region $h/J \lesssim 0.4$, the fidelity initially starts from its zero-field asymptotic values $\mathcal{F}=0.75$ and $0.8125$ for the two considered input states with $\protect\theta=\protect\pi/2$ and $\protect\pi/3$, respectively. The fidelity then exhibits a sudden increase to its maximum possible value $\mathcal{F}=1$ near the transition field $h/J \approx 0.4$ related to the magnetic-field-driven phase transition from the QAF to QFI ground state. The higher initial value of the fidelity $\mathcal{F}=0.8125$ observed for the teleportation of the less entangled input state $\protect\theta=\protect\pi/3$ may seem somewhat surprising, but this enhanced value can be attributed to the partial entanglement of the quantum channel driven into the QAF ground state that is not capable of transmitting the fully entangled input state without some disturbance. In contrast, when the quantum channel is brought into the fully entangled QFI ground state, the communication channel is free of this drawback as it can transmit quantum information effectively regardless of whether the input state is fully or partially entangled. For this reason, moderate values of the magnetic field $0.4 \lesssim h/J \lesssim 1.5$ significantly enhance the fidelity of quantum teleportation irrespective of whether the transmitted two-qubit state is fully or partially entangled. 
However, as the magnetic field is increased further, the fidelity shows a noticeable drop around the transition field $h/J \approx 1.5$ between the QFI and CPM ground states. This drop can be attributed to the fully classical nature of the CPM ground state, which is incapable of supporting entanglement. Owing to this fact, the fidelity for the quantum teleportation of both fully and partially entangled two-qubit states asymptotically reaches, after passing through a certain local minimum, the values of $\mathcal{F}=0.5$ and $0.375$, which fall below the classical threshold value $\mathcal{F}_{clas}=2/3$ indicating impossibility of quantum teleportation in this region.  In Fig.~\ref{figf}(a) and (b) we also present by broken lines typical dependencies of the average fidelity $\mathcal{F}_{av}$, which is calculated by averaging over the quantum teleportation of all possible two-qubit states. While the average fidelity $\mathcal{F}_{av}$ captures in a range of moderate magnetic fields $0.4 \lesssim h/J \lesssim 1.5$ essentially the same features as the fidelity for fully and partially entangled two-qubit states discussed previously, it exhibits distinct behaviour in the  the low- and high-field regions. Specifically, the average fidelity starts from a higher asymptotic value $\mathcal{F}_{av}=5/6$ in the low-field region and tends towards a smaller asymptotic value of $\mathcal{F}_{av}=1/3$ in the high-field region. 

\begin{figure}[t]
\includegraphics[width=0.5\textwidth]{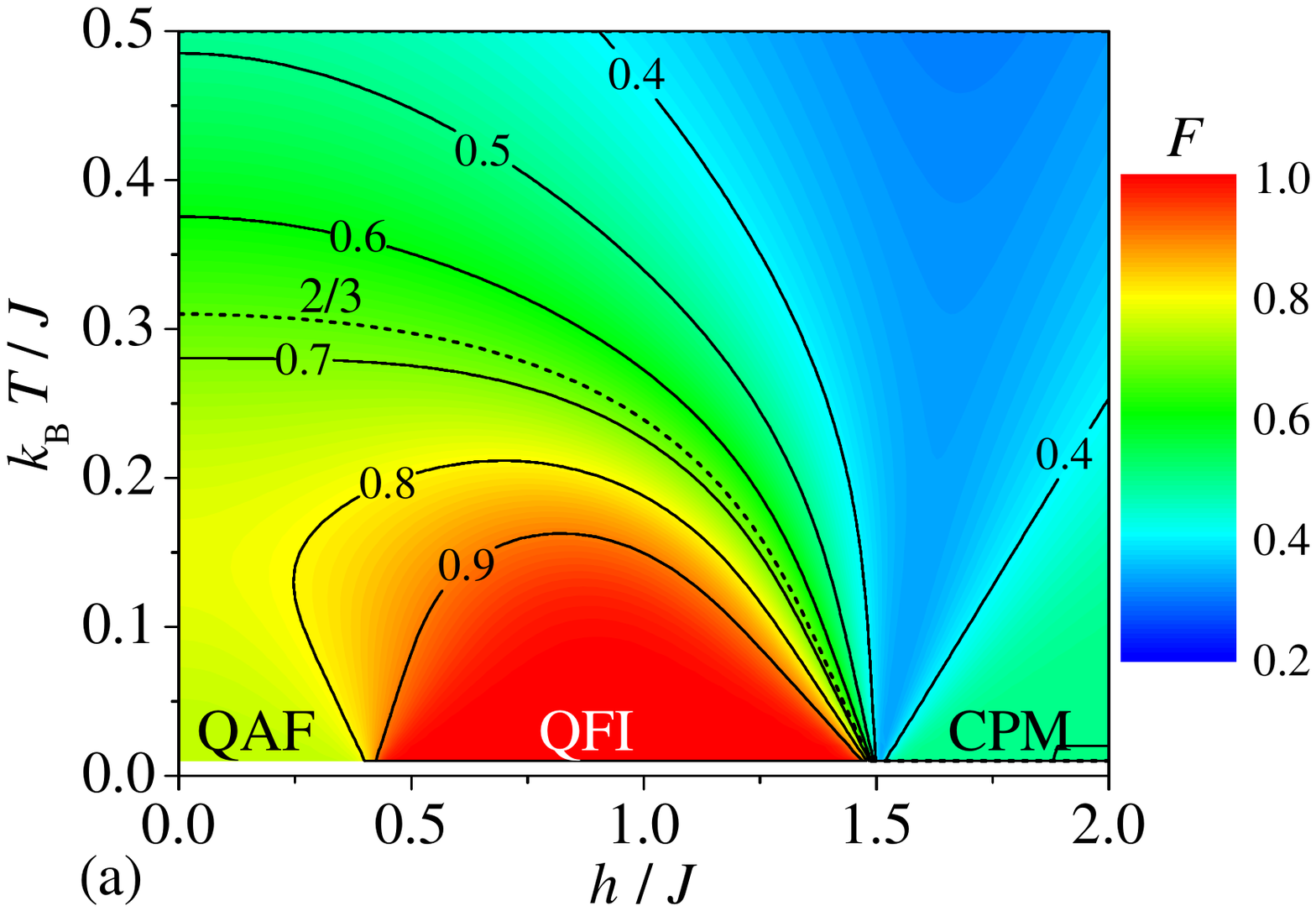}
\includegraphics[width=0.5\textwidth]{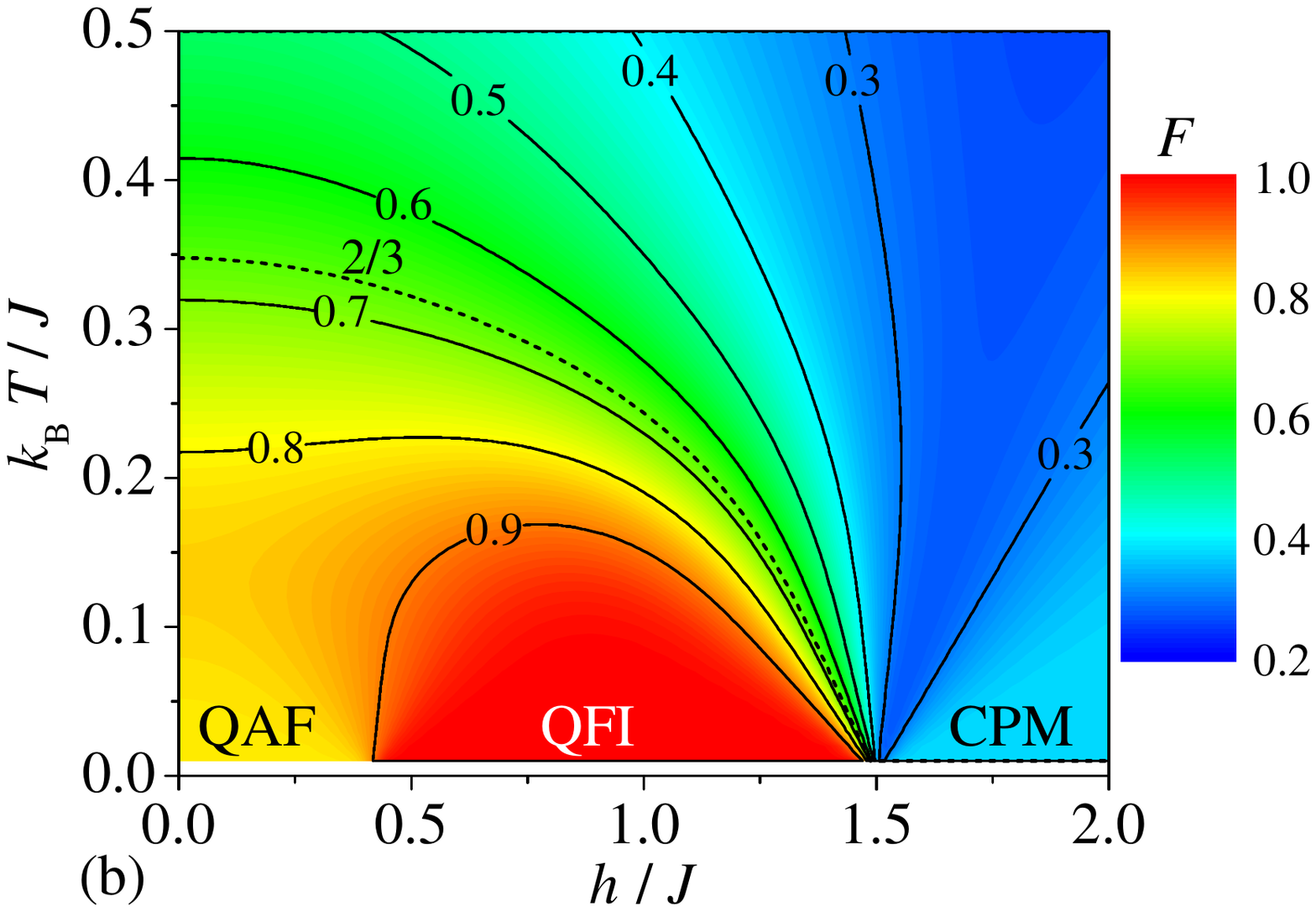} 
\includegraphics[width=0.5\textwidth]{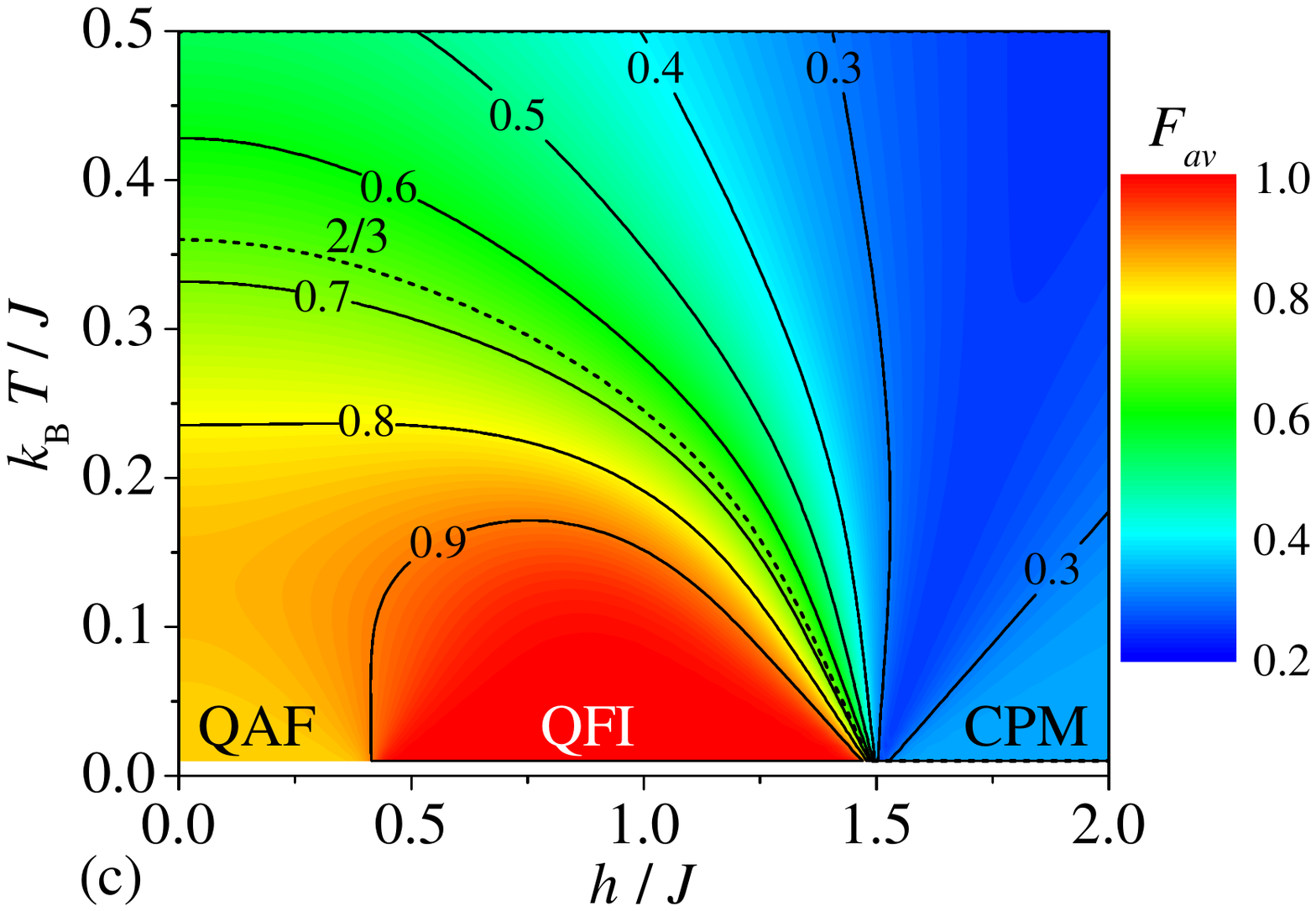} 
\vspace{-0.4cm}
\caption{(a)-(b) A density plot of the fidelity $\mathcal{F}$ for the quantum teleportation of two-qubit input state with the mixing angle $\protect\theta=\protect\pi/2$ (a) or $\protect\theta=\protect\pi/3$ (b) in the plane $h/J-k_{\rm B}T/J$ for the fixed value of the interaction ratio $J_{1}/J=1$; (c) A density plot of the average fidelity $\mathcal{F}_{av}$ of the quantum teleportation in the plane $h/J-k_{\rm B}T/J$ for the fixed value of the interaction ratio $J_{1}/J=1$. A broken line shows a contour line for the threshold value of $\mathcal{F}=2/3$ and $\mathcal{F}_{av}=2/3$.}
\label{figfd}
\end{figure}

To gain a comprehensive understanding of quantum teleportation through the quantum channel formed by two Heisenberg dimers of the spin-1/2 Ising-Heisenberg trimer chains, we finally display in Fig.~\ref{figfd}(a) and (b) the density plot of the fidelity $\mathcal{F}$ in the magnetic field $h/J$ versus temperature $k_{\rm B}T/J$ plane for two entangled two-qubit states with  mixing angles $\theta =\pi/2$ and $\pi/3$ by considering a fixed value of the interaction ratio $J_{1}/J=1$. It is clear from these figures that the high-fidelity quantum teleportation with $\mathcal{F} \gtrsim 0.9$ is confined to a dome-like parameter region, which can be allocated to moderately strong magnetic fields $0.4 \lesssim h/J \lesssim 1.5$ and sufficiently low temperatures $k_{\rm B}T/J \lesssim 0.15$. A contour line for the specific value of the fidelity $\mathcal{F}=2/3$, which is shown in Fig.~\ref{figfd}(a) and (b) by a broken line, sets the upper bound for transmitting quantum information via quantum teleportation. Fig.~\ref{figfd}(a) and (b) also demonstrate that the less entangled input state with the mixing angle $\theta =\pi/3$ can be transmitted more efficiently through this quantum channel than the fully entangled state with the mixing angle $\theta =\pi/2$ as the respective fidelity values are higher and reach the classical threshold $\mathcal{F}_{clas}=2/3$ just at a higher temperature (c.f. $k_{\rm B}T/J \approx 0.35$ for $\theta =\pi/3$ with $k_{\rm B}T/J \approx 0.31$ for $\theta =\pi/2$). 

We have thus convincingly evidenced that the quantum channel formed by two Heisenberg dimers of the spin-1/2 Ising-Heisenberg trimer chains serves as an efficient platform for realizing quantum teleportation of arbitrarily entangled two-qubit state. Finally, the average fidelity is displayed in Fig.~\ref{figfd}(c) in the magnetic field versus temperature plane for a fixed value of the interaction ratio $J_{1}/J=1$. Generally, the average fidelity $\mathcal{F}_{av}$ qualitatively exhibits almost the same features as the fidelity $\mathcal{F}$ especially when the density plot of the average fidelity $\mathcal{F}_{av}$ is compared to that one of the fidelity for the partially entangled input state with the mixing angle $\theta = \pi /3$. The key quantitative difference arises in the low-field and low-temperature parameter region, where the average fidelity reaches a somewhat higher value compared to the fidelity acquired for the quantum teleportation of both considered two-qubit input states with the mixing angle $\theta =\pi/2$ and $\pi/3$. 

\section{Copper-based trimer chains as resource for quantum channel}
\label{resexp}

\begin{figure}[t!]
\begin{center}
\includegraphics[width=0.49\textwidth]{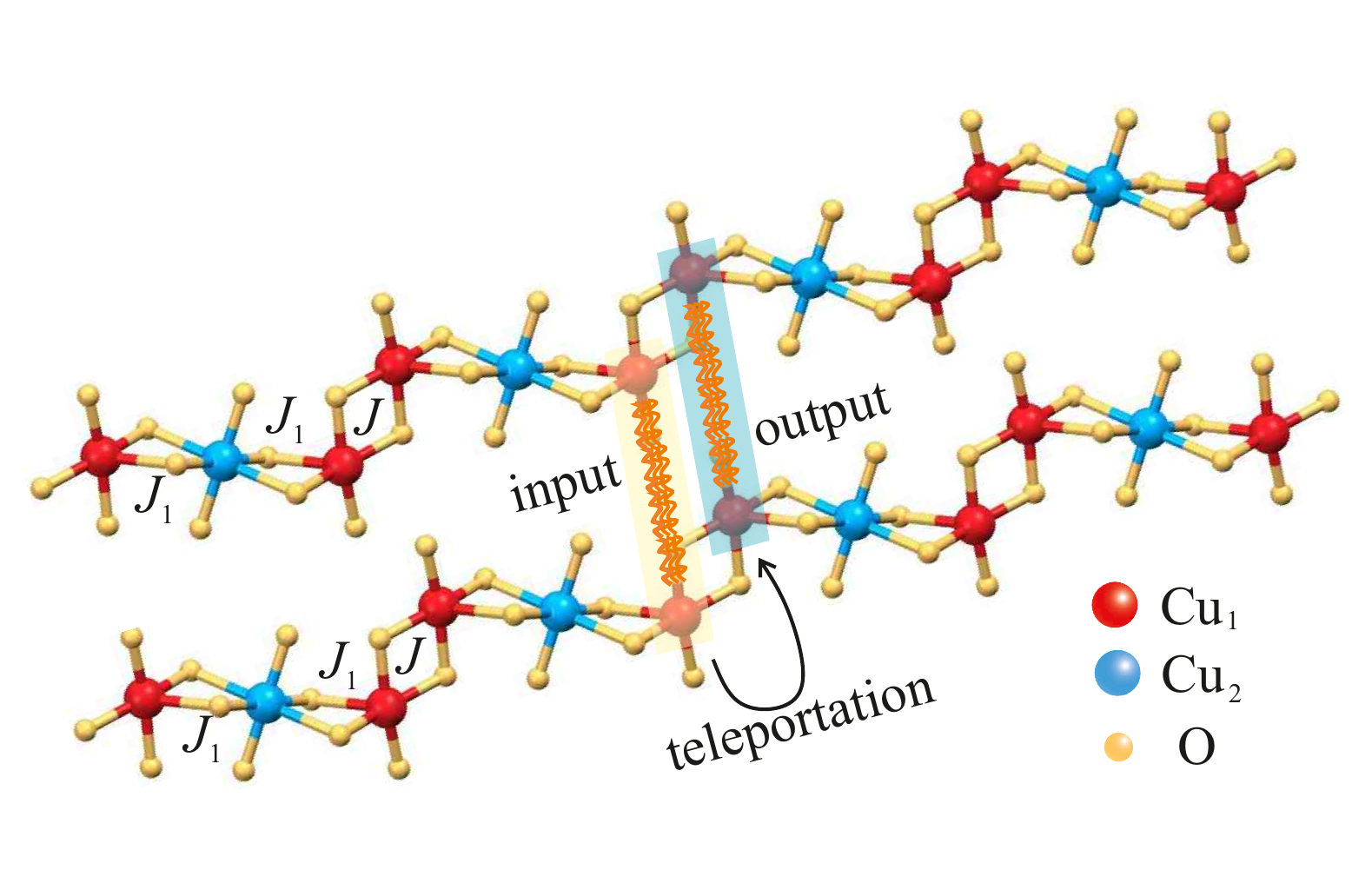}
\end{center}
\vspace{-0.7cm}
\caption{A part of the crystal structure of two polymeric trimer chains Cu$_3$(P$_2$O$_6$OH)$_2$ serving as resource for the quantum communication channel. Red and blue spheres denote lattice positions of two crystallographically inequivalent spin-1/2 Cu$^{2+}$ magnetic ions denoted as Cu$_1$ and Cu$_2$, which are coupled within the polymeric chains Cu$_3$(P$_2$O$_6$OH)$_2$ through two different coupling constants $J$ and $J_1$ providing a trimerized pattern of the coupling constants $J_1-J_1-J$. A quantum teleportation of a two-qubit input state is considered within the quantum channel, which is composed from the Cu$_1$-Cu$_1$ dimeric units of two polymeric trimer chains Cu$_3$(P$_2$O$_6$OH)$_2$.}
\label{latticeexp}
\end{figure}

In this section, we will explore quantum teleportation through a quantum communication channel composed of two polymeric trimer chains Cu$_3$(P$_2$O$_6$OH)$_2$ \cite{HASE1,HASE2,HASE3,KONG2015,HASE4} as illustrated in Fig. \ref{latticeexp}. The magnetic backbone of the polymeric compound Cu$_3$(P$_2$O$_6$OH)$_2$ is formed by a regular alternation of the monomer Cu$^{2+}$ magnetic ions labeled as Cu$_2$ with the dimeric Cu$^{2+}$-Cu$^{2+}$ entities labeled as Cu$_1$-Cu$_1$. Theoretical modeling of magnetization and susceptibility data over a wide range of temperatures and magnetic fields indicates that the exchange interaction inside of the dimeric Cu$_1$-Cu$_1$ units is described by a dominant coupling constant $J/k_{\rm B} \approx 103$~K, while the exchange interaction between the monomeric Cu$_2$ and dimeric Cu$_1$ magnetic ions is characterized by a significantly weaker coupling constant $J_1/k_{\rm B} \approx 30$~K \cite{VER21}. 

Although the polymeric compound Cu$_3$(P$_2$O$_6$OH)$_2$ essentially represents an experimental realization of a spin-1/2 Heisenberg trimer chain, inelastic neutron diffraction experiments conducted at a magnetic field of 6~T and a temperature of 1.6~K reveal a highly polarized nature of the monomer spins $m$(Cu$_2$) $\approx 0.43 \mu_{\rm B}$ and an almost unpolarized nature of the dimeric spins $m$(Cu$_1$) $ \approx 0.01 \mu_{\rm B}$ \cite{HASE4}. This observation suggest that the spin-1/2 Ising-Heisenberg trimer chain provides a reasonable approximation to the spin-1/2 Heisenberg trimer chain due to the relatively small quantum fluctuations affecting the monomeric spins, which diminish further upon increasing the magnetic field. The monomeric spins indeed become fully polarized at the onset of the intermediate one-third magnetization plateau emerging approximately at the magnetic field of 12~T \cite{HASE1,KONG2015}. Beyond this magnetic field, the simplified spin-1/2 Ising-Heisenberg trimer chain provides an accurate representation of the magnetic compound Cu$_3$(P$_2$O$_6$OH)$_2$ as quantum fluctuations at the monomeric spins are completely suppressed. This approach has already been successfully applied for a theoretical modeling of magnetic properties of the polymeric compound Cu(3-Chloropyridine)$_2$(N$_3$)$_ 2$, which affords an experimental realization of a spin-1/2 Heisenberg tetramer chain \cite{str04,str05}.

\begin{figure}[t]
\includegraphics[width=0.5\textwidth]{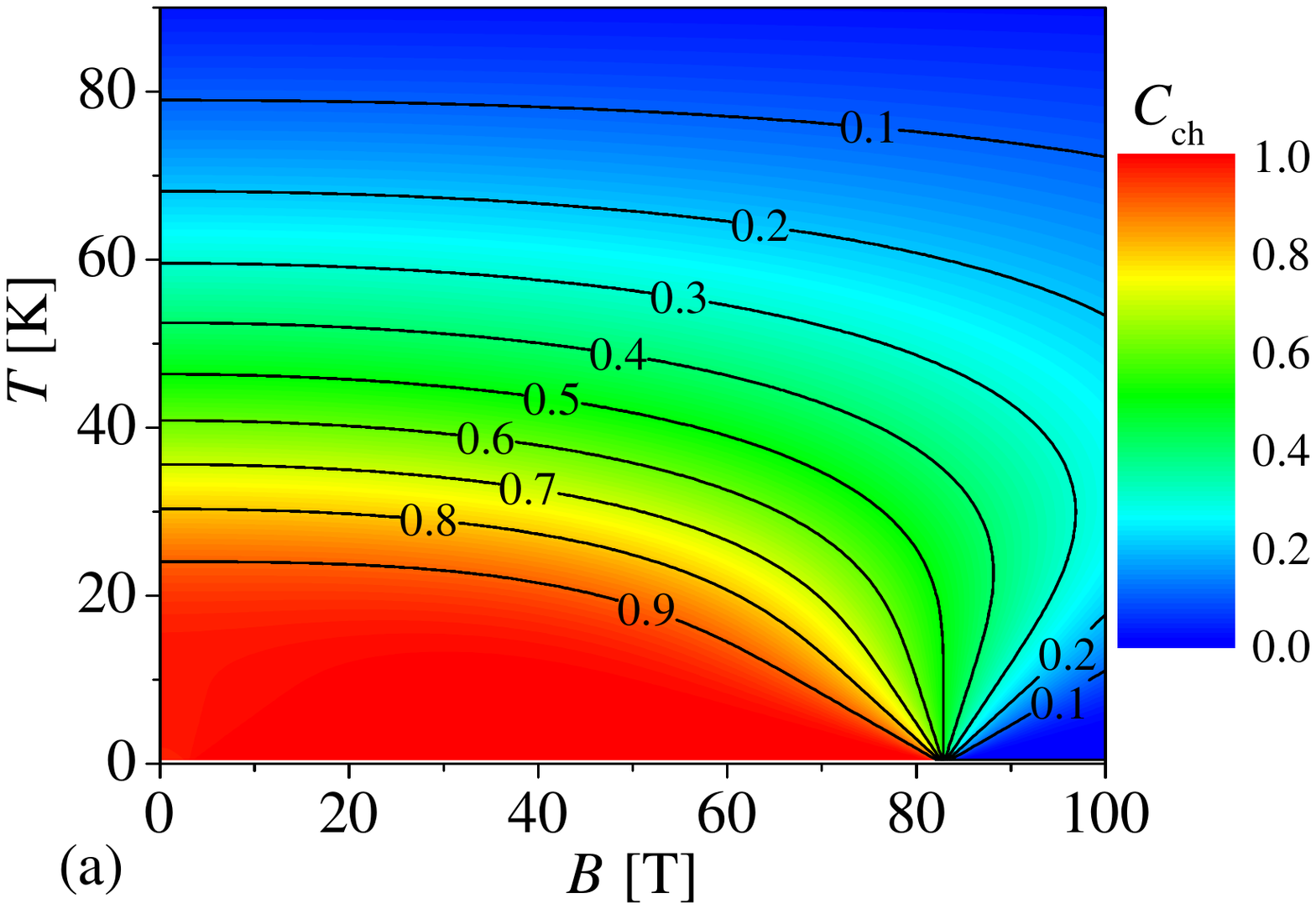}
\includegraphics[width=0.5\textwidth]{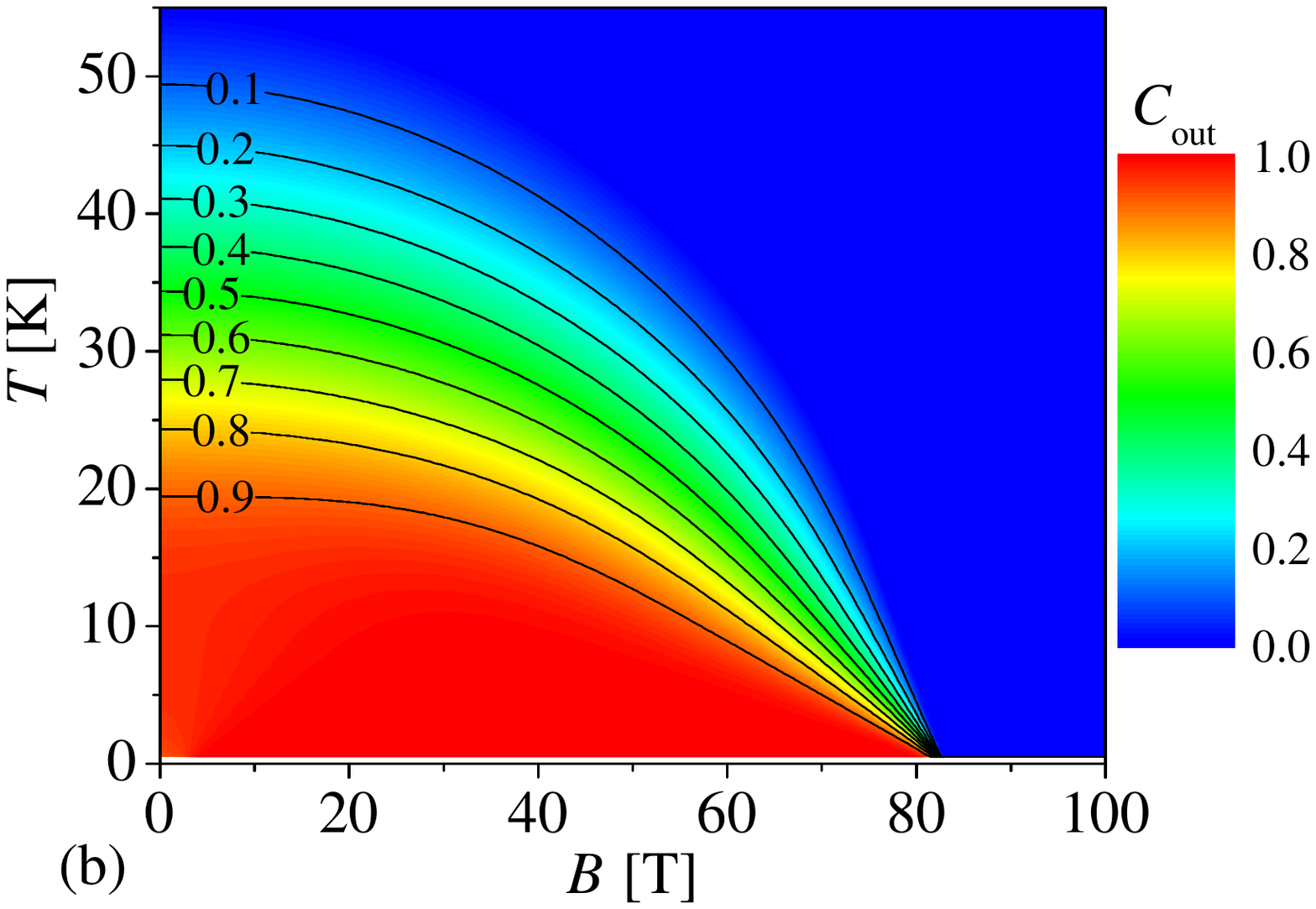}
\includegraphics[width=0.5\textwidth]{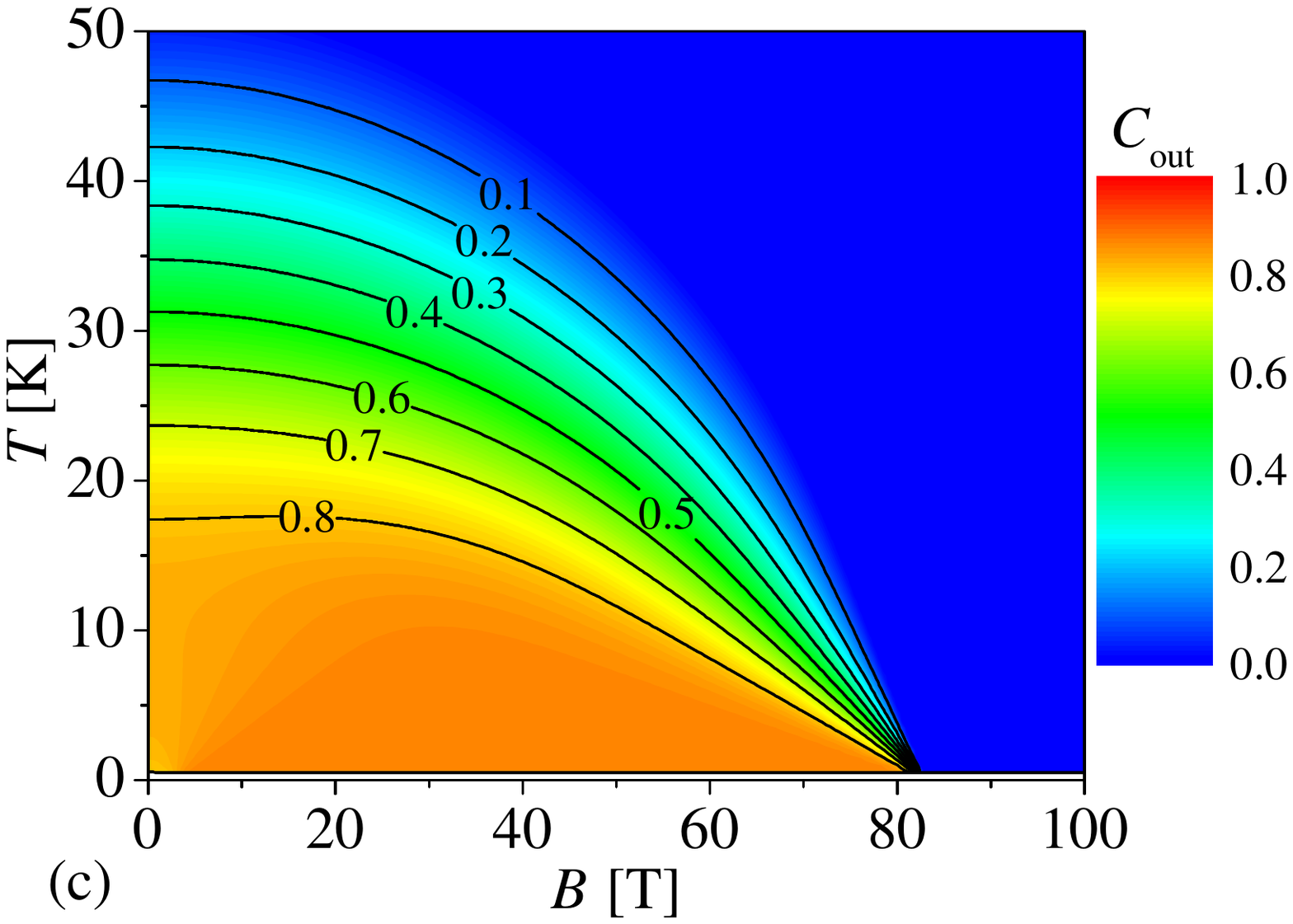}
\vspace{-0.4cm}
\caption{(a) A density plot of the concurrence of the quantum channel $\mathcal{C}_{ch}$ composed from two polymeric trimer chains Cu$_3$(P$_2$O$_6$OH)$_2$ in the magnetic field versus temperature plane. The polymeric magnetic compounds are modeled by spin-1/2 Ising-Heisenberg trimer chains with the Land\'e g-factor $g=2.12$ and two different coupling constants $J/k_{\rm B} = 103$~K and $J_1/k_{\rm B} = 30$~K; (b)-(c) Density plots of the concurrence of the output state $\mathcal{C}_{out}$ in the magnetic field versus temperature plane by implementing this quantum communication channel for two different input states with the mixing angle: 
(b) $\protect\theta=\protect\pi/2$, (c) $\protect\theta=\protect\pi/3$.}
\label{figce}
\end{figure}

\begin{figure}[t]
\includegraphics[width=0.5\textwidth]{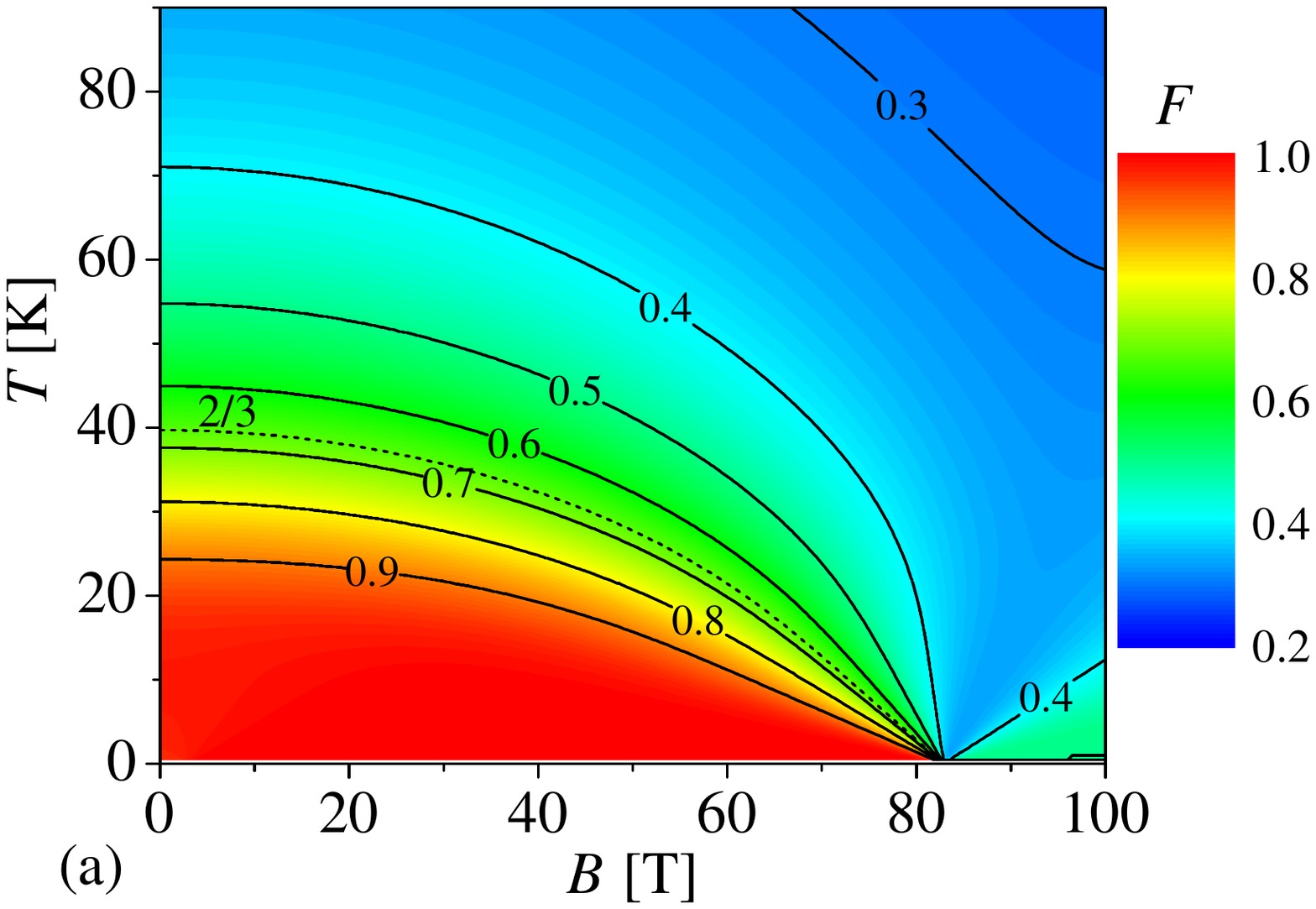}
\includegraphics[width=0.5\textwidth]{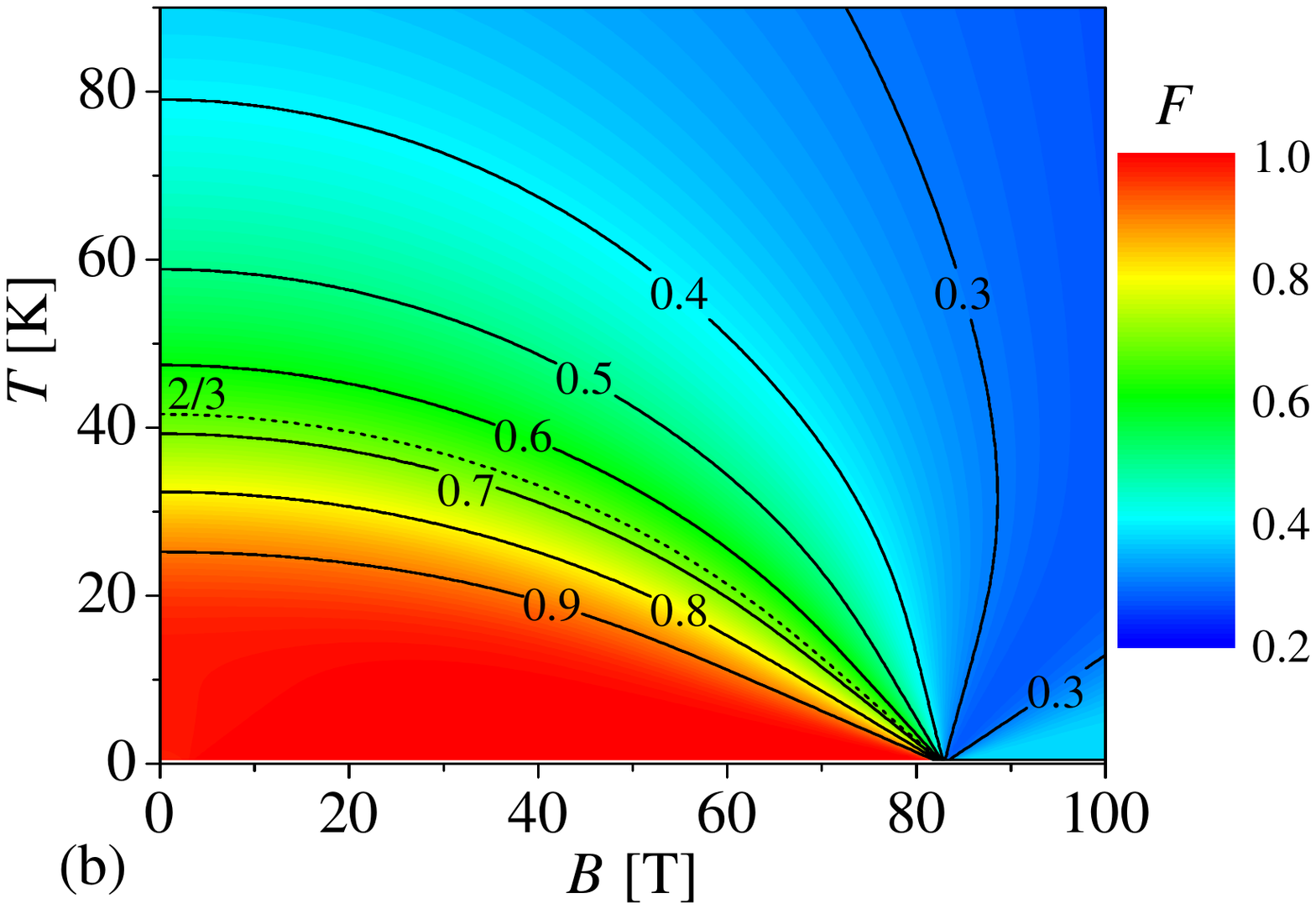} 
\includegraphics[width=0.5\textwidth]{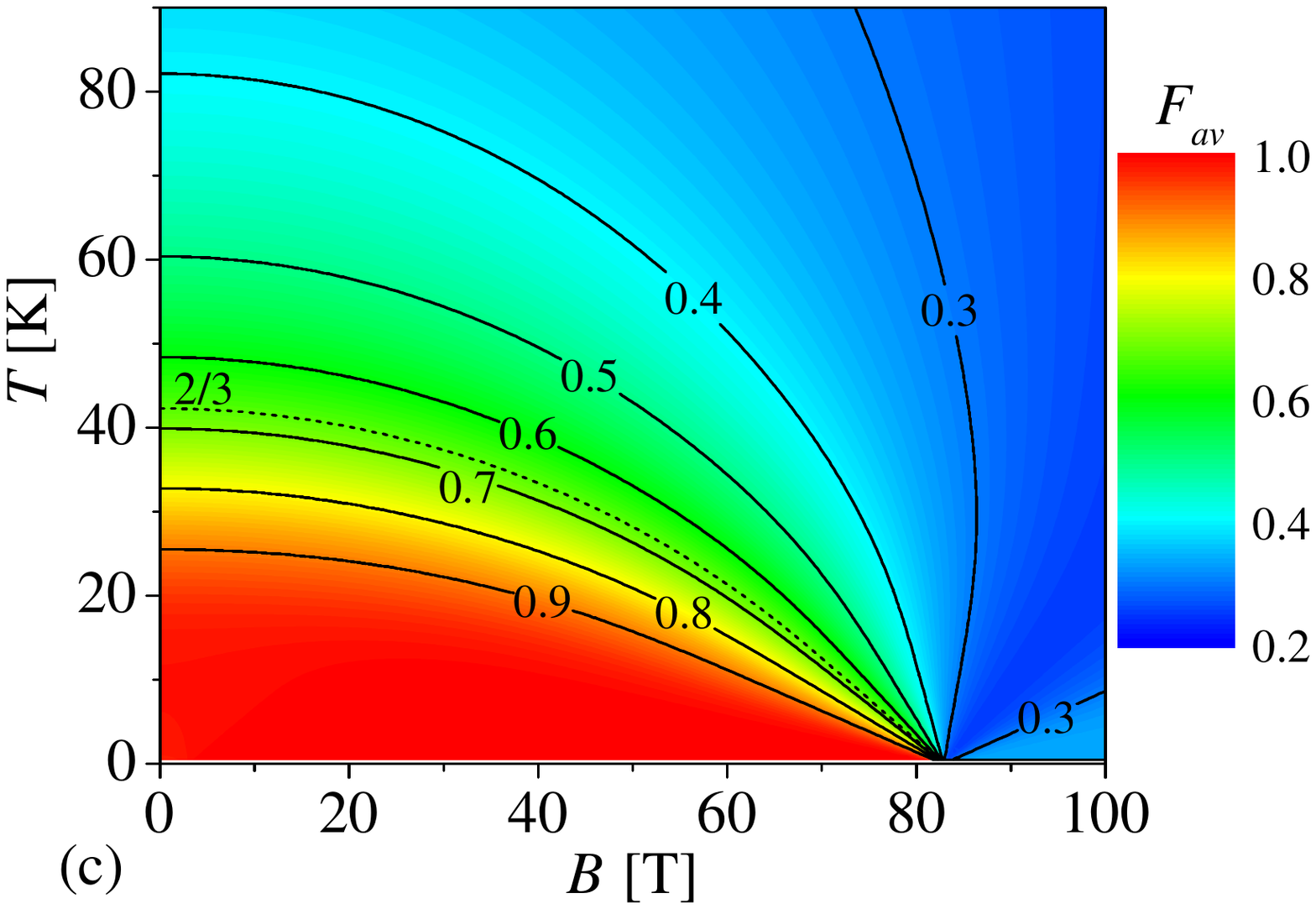} 
\vspace{-0.4cm}
\caption{(a)-(b) A density plot of the fidelity $\mathcal{F}$ for the quantum teleportation of two-qubit input state with the mixing angle $\protect\theta=\protect\pi/2$ (a) or $\protect\theta=\protect\pi/3$ (b) in the magnetic field versus temperature plane; (c) A density plot of the average fidelity $\mathcal{F}_{av}$ of the quantum teleportation of an arbitrary two-qubit input state in the magnetic field versus temperature plane. The quantum communication channel is formed by the polymeric trimer chains Cu$_3$(P$_2$O$_6$OH)$_2$ described by spin-1/2 Ising-Heisenberg trimer chains with the Land\'e g-factor $g=2.12$ and two different coupling constants $J/k_{\rm B} = 103$~K and $J_1/k_{\rm B} = 30$~K. A broken line shows a contour line for the threshold value of $\mathcal{F}=2/3$ and $\mathcal{F}_{av}=2/3$.}
\label{figef}
\end{figure}

The concurrence of the quantum channel $\mathcal{C}_{ch}$ involving two dimeric Cu$_1$-Cu$_1$ units embedded in independent polymeric trimer chains Cu$_3$(P$_2$O$_6$OH)$_2$ is plotted in Fig.~\ref{figef}(a). For the respective theoretical modeling we have adapted the spin-1/2 Ising-Heisenberg trimer chains with the average value of Land\'e g-factor $g=2.12$ for all Cu$^{2+}$ magnetic ions as determined by electron spin resonance measurements \cite{HASE1,HASE2} and the coupling constants $J/k_{\rm B} = 103$~K and $J_1/k_{\rm B} = 30$~K obtained from the best theoretical fit of magnetization and susceptibility data \cite{VER21}. The concurrence of the quantum channel driven towards the QAF and QFI ground state becomes nearly identical $C_{ch}^{QAF} \approx 0.96$ vs. $C_{ch}^{QFI} = 1$ due to a relatively small value of the interaction ratio $J_1/J \approx 0.3$. This fact consequently precludes to identify clear signatures of the magnetic-field-driven transition between the QAF and QFI phases unlike the particular case with an identical strength of the coupling constant $J_1/J = 1$ discussed previously. It is evident from Fig.~\ref{figef}(a) that the strong bipartite entanglement of the dimeric Cu$_1$-Cu$_1$ units of the quantum channel driven towards QAF and QFI phases consequently persists up to relatively high temperatures $T \approx 80$~K and extremely high magnetic fields $B \approx 80$~T. The breakdown of entanglement of the quantum channel observed at sufficiently low temperatures coincides with a magnetic-field-driven transition from the QFI ground state to the fully polarized CPM ground state. Notably, the respective threshold value of the magnetic field $B \approx 80$~T is in a perfect agreement with the spin gap of 9.8 meV observed for Cu$_3$(P$_2$O$_6$OH)$_2$ in inelastic neutron scattering experiments \cite{HASE2,HASE3}.

Furthermore, the concurrence of the output state $\mathcal{C}_{out}$ transmitted through the quantum communication channel is displayed in Fig.~\ref{figef}(b)-(c) for two different input states characterized by the mixing angles $\protect\theta=\protect\pi/2$ and $\protect\theta=\protect\pi/3$. While the threshold magnetic field marking the breakdown of bipartite entanglement in the output state remains practically unchanged, the sudden-death temperature associated with disappearance of the bipartite thermal entanglement is progressively reduced to much lower temperatures $T \approx 50$~K. It also follows from Fig.~\ref{figef}(b)-(c) that the concurrence of the output state readout from the quantum channel evidently has one additional distinct feature from the concurrence of the quantum channel itself. While a thermally-assisted rise of the concurrence of the quantum channel $\mathcal{C}_{ch}$ is possible even at high enough magnetic fields favoring the CPM ground state, the concurrence of the output state $\mathcal{C}_{out}$ may not be thermally induced above the CPM ground state. 

Finally, let us conclude our discussion with a detailed analysis of the density plots of fidelity and average fidelity in the magnetic field versus temperature plane (see Fig. \ref{figef}), which were calculated for the quantum channel composed of two independent polymeric trimer chains Cu$_3$(P$_2$O$_6$OH)$_2$ using spin-1/2 Ising-Heisenberg trimer chains with the Land\'e g-factor $g=2.12$ and two different coupling constants $J/k_{\rm B} = 103$~K and $J_1/k_{\rm B} = 30$~K. The parameter region with sufficiently high values of the fidelity $\mathcal{F}>2/3$ and average fidelity $\mathcal{F}_{av}>2/3$ determine the conditions, under which the quantum communication channel outperforms an efficiency of any classical communication channel. It directly follows from the comparison of Fig. \ref{figef}(a)-(c) that the quality of quantum teleportation is nearly identical for the transmission of fully (a), partially (b) or arbitrarily (c) entangled two-qubit pure state. Specifically, the quantum communication channel formed by the two polymeric trimer chains Cu$_3$(P$_2$O$_6$OH)$_2$ enables efficient quantum teleportation on assumption that temperature remains below $T \lesssim 40$~K and the magnetic field stays below $B \lesssim 80$~T.

\section{Concluding remarks}
\label{conclusion}

In the present paper, we have rigorously investigated the quantum teleportation of a two-qubit pure state through a quantum channel, which is composed of two Heisenberg dimers embedded in spin-1/2 Ising-Heisenberg trimer chains. Using the transfer-matrix formalism, we derived exact analytical results for the concurrence of the quantum channel, the concurrence of the output state, the fidelity, and the average fidelity, which can be regarded as key metrics to assess the strength of quantum entanglement and the precision of quantum teleportation protocol through this quantum channel. It has been demonstrated that efficient quantum teleportation surpassing the capabilities of classical communication can be reached within this quantum channel at sufficiently low temperatures and magnetic fields.

We have convincingly evidenced that the magnetic field may significantly enhance the overall efficiency of the quantum teleportation protocol what is in contrast with general expectations. Enhancing fidelity in the quantum teleportation due to the magnetic field arises from the strengthening of bipartite entanglement within the quantum channel as the Heisenberg dimers of spin-1/2 Ising-Heisenberg trimer chains are driven from a partially entangled singlet-dimer-like state in the QAF phase to a perfectly entangled singlet-dimer state in the QFI phase. This unexpected finding holds true regardless of whether the transmitted input state is fully or only partially entangled. At sufficiently low magnetic fields favoring the QAF phase it is even possible to achieve higher fidelity for the quantum teleportation of a partially entangled input state compared to that of a fully entangled input state within this quantum channel.

Last but not least, it has been proposed that the quantum communication channel formed by the Heisenberg dimers of two independent spin-1/2 Ising-Heisenberg trimer chains can be applied for a theoretical modeling of the quantum teleportation realized through two polymeric trimer chains Cu$_3$(P$_2$O$_6$OH)$_2$ \cite{HASE1,HASE2,HASE3,KONG2015,HASE4}. It has been found that the bipartite entanglement of this quantum channel persists up to extremely high magnetic fields $B \lesssim 80$~T and moderately high temperatures $T \lesssim 80$~K, whereas the bipartite entanglement of the output state readout from the quantum channel is detectable up to slightly smaller temperatures $T \lesssim 50$~K though the resistance against the magnetic field is fully preserved. Most intriguingly, the fidelity and average fidelity for the quantum teleportation of an entangled two-qubit input state is surpassing the efficiency of any classical communication channel up to moderately high temperatures $T \lesssim 40$~K and extremely high magnetic fields $B \lesssim 80$~T. This suggests that the quantum channel formed by two polymeric trimer chains Cu$_3$(P$_2$O$_6$OH)$_2$ holds great promise for reliably transmitting  quantum information using this quantum teleportation protocol. Apart from the aforementioned quantum spin chain, a few other polymeric copper-based compounds [Cu$_3$(3,5-dimethylpyrazole)$_2$($\mu$-benzoate)$_4$(benzoate)$_2$] \cite{deka06}, [Cu$_3$(3,5-dimethylpyrazole)$_2$(acetate)$_2$(4-methylbenzoate)$_4$] \cite{sarma}, [Cu$_3$(tris(2-carboxyethyl)isocyanurate)$_2$(H$_2$O)$_2$] \cite{li11}, [Cu$_3$(2,6-di-imidazol-1-yl-pyridine)(acetate)$_6$(CH$_3$OH)] \cite{zhang}, [Cu$_3$(4-aminobutyric acid)$_4$Cl$_4$(H$_2$O)$_2$](ClO$_4$)$_2$ \cite{bouh17}, [Cu$_3$(acetate)$_6$(imidazole)$_2$] \cite{machado18} may also serve as efficient platforms for the quantum teleportation of entangled two-qubit states through quantum channels consisting of two trimeric quantum spin chains. 

\begin{acknowledgments}
This work was financially supported by The Ministry of Education, Research, Development and Youth of the Slovak Republic under the grant number VEGA 1/0695/23 
and by Slovak Research and Development Agency under the contract number APVV-20-0150. 
\end{acknowledgments}

\begin{appendices}
\section{}
\label{app}
The coefficients $F_{ij}(x)$ entering into individual elements of the reduced density matrix (\ref{dme}) are given by the following expressions:
\begin{widetext}
\begin{eqnarray}
F_{11} (x) &=& \frac{\exp \left(-\frac{\beta J}{4} - \frac{\beta J_1}{2} x + \beta h \right)}{2 \exp \left(-\frac{\beta J}{4}\right) \cosh \left(\frac{\beta J_1}{2} x - \beta h \right) 
+ 2 \exp \left(\frac{\beta J}{4}\right) \cosh \left(\frac{\beta}{2} \sqrt{J_1^2 (1 - x^2) + J^2} \right)}, \nonumber \\
F_{22} (x) &=& F_{33} (x) = \frac{\exp \left(\frac{\beta J}{4}\right) \cosh \left(\frac{\beta}{2} \sqrt{J_1^2 (1 - x^2) + J^2} \right)}{2 \exp \left(-\frac{\beta J}{4}\right) \cosh \left(\frac{\beta J_1}{2} x - \beta h \right) + 2 \exp \left(\frac{\beta J}{4}\right) \cosh \left(\frac{\beta}{2} \sqrt{J_1^2 (1 - x^2) + J^2} \right)}, \nonumber \\
F_{44} (x) &=& \frac{\exp \left(-\frac{\beta J}{4} + \frac{\beta J_1}{2} x - \beta h \right)}{2 \exp \left(-\frac{\beta J}{4}\right) \cosh \left(\frac{\beta J_1}{2} x - \beta h \right) 
+ 2 \exp \left(\frac{\beta J}{4}\right) \cosh \left(\frac{\beta}{2} \sqrt{J_1^2 (1 - x^2) + J^2} \right)}, \nonumber \\
F_{23} (x) &=& F_{32} (x) = - \frac{J}{\sqrt{J_1^2 (1-x^2) + J^2}} \frac{\exp \left(\frac{\beta J}{4}\right) \sinh \left(\frac{\beta}{2} \sqrt{J_1^2 (1 - x^2) + J^2} \right)}{2 \exp \left(-\frac{\beta J}{4}\right) \cosh \left(\frac{\beta J_1}{2} x - \beta h \right) + 2 \exp \left(\frac{\beta J}{4}\right) \cosh \left(\frac{\beta}{2} \sqrt{J_1^2 (1 - x^2) + J^2} \right)}. \nonumber 
\end{eqnarray}
\end{widetext}
\end{appendices}

\end{document}